\PassOptionsToPackage{numbers,sort&compress}{natbib}
\documentclass[final,5p,times,twocolumn]{elsarticle}

\usepackage{amssymb}
\usepackage{amsmath}
\usepackage{graphicx}
\usepackage{xcolor}
\usepackage{bm}        % for math
\usepackage{graphicx}
\usepackage{slashed}
\setcitestyle{numbers,sort&compress}
\usepackage{hyperref}

\begin{document}

\begin{frontmatter}



\title{Search for dark photon and dark matter signatures around electron-positron colliders}


\author[addr1,addr3]{Xin Chen}
\ead{xin.chen@cern.ch}
\author[addr1,addr3]{Zhen Hu}
\ead{Zhen.Hu@cern.ch}
\author[addr5]{Yongcheng Wu}
\ead{ycwu@physics.carleton.ca}
\author[addr6]{Kai Yi}
%\ead{06272@njnu.edu.cn}

\address[addr1]{Department of Physics, Tsinghua University, Beijing 100084, China}
%\address[addr2]{Collaborative Innovation Center of Quantum Matter, Beijing 100084, China}
\address[addr3]{Center for High Energy Physics, Tsinghua University, Beijing 100084, China}
\address[addr5]{Ottawa-Carleton Institute for Physics, Carleton University, Ottawa, Ontario K1S 5B6, Canada}
\address[addr6]{School of Physics and Technology, Nanjing Normal University, 1 Wenyuan Road, 210046, China}

\begin{abstract}

The search for a dark photon produced at $e^{+}e^{-}$ colliders which subsequently
decays into inelastic dark matter particles, is discussed. The heavier
dark matter decays into a pair of visible charged particles and a lighter
dark matter particle after traveling some distance. The visible decay products
can be recorded by a dark matter detector made of emulsions and gas detectors, placed
near the main $e^{+}e^{-}$ detector. This setup can not only explore new
parameter regions not reached before, but also re-open some regions thought
to be excluded by previous experimental data. The physics potential for
such a detector around BESIII and Belle II is presented.

\end{abstract}


\end{frontmatter}

\section{Introduction}

The dark matter (DM) is well motivated by astrophysical observations, and
it is believed to interact weakly with the Standard Model (SM) particles
through some mediators like the dark photon. A number of fixed-target experiments,
as well as collider-based ones, have searched for a dark photon, where
it is produced from charged particles with a tiny coupling strength. The
dark photon can decay back into charged SM particles, after flying some
distance due to the small coupling and hence small decay width, which is
the model of many fixed-target experiments (see \cite{BeamDump} and references
therein). When the dark photon mass is large and the coupling is not so
small, it can decay promptly into a pair of charged SM particles or neutral
DM ones, which has been studied at the electron-positron colliders such
as BaBar \cite{BaBar_Atoll,BaBar_AtoInv}.

With ever more stringent limits obtained from direct DM detection experiments
\cite{XENON1T}, a WIMP-like DM particle with mass $\gtrsim $ 10 GeV is
being severely constrained, and a lot of attention is being paid to light
DM searches \cite{L_DM_1,L_DM_2,L_DM_3,L_DM_4}. On the other hand, Cosmic
Microwave Background (CMB) anisotropy result \cite{RD} excluded a DM particle
mass $\lesssim $ 10 GeV for the WIMP $s$-wave annihilation scenario consistent
with the relic DM abundance. The tension can be relaxed with an inelastic
DM model \cite{Excited_DM_1}. One of the two DM particles is in a heavy
mass eigenstate, so it can further decay into a ground state DM and a charged
SM particle pair. By detecting the charged pair, the existence of a DM
sector with a dark photon mediator can be inferred. We here propose a new
experimental setup to search for a dark photon that promptly decays into
such a pair of DM particles near the $e^{+}e^{-}$ colliders. The DM detector
can be placed in the forward or central region where the heavy ``excited''
DM particles are more abundantly produced, but are difficult to be detected
by the main $e^{+}e^{-}$ detector.

%s2 #&#
\section{Theoretical framework}

At low energy, we assume an effective toy model with a dark photon mediator
$A'$ and complex scalar $\phi $ as the DM particle. The relevant Lagrangian
reads
%
%e1 #&#
\begin{align}
\label{equ:lagr}
\mathcal{L} = & \mathcal{L}_{SM} - \frac{1}{4} A^{\prime }_{\mu \nu }A^{
\prime \mu \nu } + \frac{1}{2} m^{2}_{A'} A^{\prime }_{\mu }A^{\prime \mu } -
\epsilon e Q_{f} A^{\prime }_{\mu }\bar{f}\gamma ^{\mu }f +
\nonumber
\\
& \left ( D_{\mu }\phi \right )^{*} D^{\mu }\phi - \mu ^{2} \phi ^{*}
\phi - \frac{1}{2}\rho ^{2}\left (\phi \phi + \phi ^{*} \phi ^{*}
\right ),
\end{align}
where $A'$ is the dark photon, $f$ is the SM fermion with charge
$Q_{f}$, $\epsilon $ is the mixing parameter between the $U(1)_{Y}$ and
$U(1)_{D}$ gauge fields,
$D_{\mu }=\partial _{\mu }+ i g_{D} A^{\prime }_{\mu }$ is the covariant derivative
with dark coupling parameter $g_{D}$,
$\phi =(\phi _{1}+i\phi _{2})/\sqrt{2}$ is the complex scalar DM. The last
term in Eq.~\ref{equ:lagr} is a ``Majorana''-like mass term that violates
the $U(1)_{D}$ symmetry. Substituting $\phi _{1,2}$ for $\phi $, we have
%
%e2 #&#
\begin{align}
\label{equ:lagr2}
\mathcal{L} \supset & \frac{1}{2} \partial _{\mu }\phi _{1} \partial ^{\mu }\phi _{1} + \frac{1}{2} \partial _{\mu }\phi _{2} \partial ^{\mu
}\phi _{2} - \frac{1}{2} \left ( \mu ^{2} +\rho ^{2} \right ) \phi _{1}^{2}
\nonumber
\\
& - \frac{1}{2} \left ( \mu ^{2} -\rho ^{2} \right ) \phi _{2}^{2} - g_{D}
A^{\prime }_{\mu }\left (\phi _{2}\partial ^{\mu }\phi _{1} - \phi _{1}
\partial ^{\mu }\phi _{2} \right )
\nonumber
\\
& + \frac{1}{2} g_{D}^{2} A^{\prime }_{\mu }A^{\prime \mu } \left ( \phi _{1}^{2}
+ \phi _{2}^{2} \right ).
\end{align}
It is evident that the presence of the small $\rho $-term in Eq.~\ref{equ:lagr} causes a mass splitting between the two real scalars
$\phi _{1}$ and $\phi _{2}$, with mass
$m_{1}=\sqrt{\mu ^{2} +\rho ^{2}}$ and
$m_{2}=\sqrt{\mu ^{2} -\rho ^{2}}$, respectively. It is also possible to
achieve a similar mass splitting for a fermionic DM by the presence of
Majorana mass terms \cite{Excited_DM}, but the scalar model we are studying
will be general enough to cover similar kinematics.

%s3 #&#
\section{DM relic density}
\label{sec:RD}

The inelastic DM model we investigate here has several features to relax
the tension between a light DM mass and the CMB constraint. The mass relation
$m_{A'} > m_{1}+m_{2}$ ensures that
$\phi _{2} \phi _{2} \to A' A' \to \text{SM}$ is closed, and the dominant
coannihilation channel
$\phi _{1} \phi _{2} \to A^{\prime *} \to \text{SM}$ will be no longer active
in the recombination era due to the depletion of $\phi _{1}$, which therefore
does not affect CMB much. The Lagrangian form in Eq.~\ref{equ:lagr2} means
that the elastic scattering between $\phi _{2}$ and SM particles are loop
suppressed, and thus can evade the direct DM detection via nuclear recoil
(apart from the fact that $\phi _{2}$ is light). This type of dark photon
decaying into inelastic light DM particles can be searched in
$e^{+}e^{-}$ collider experiments \cite{Excited_DM}, such as Belle II or
BESIII, with distinct signatures such as a mono-photon alone
\cite{BaBar_AtoInv}, or a mono-photon plus a pair of $e^{+}e^{-}$ with
a displaced vertex \cite{monophoton_plus_ee}. The inelastic DM has also
been discussed in the context of hadron colliders covering a dark photon
mass ${\gtrsim}10$~GeV \cite{Inelastic_had}, and fixed-target experiments
suitable for a DM mass $\lesssim $1 GeV \cite{BeamDump_2}.

It is assumed that the dark and SM matter were in thermal equilibrium in
the early universe. When the DM annihilation rate into SM particles falls
below the Hubble expansion rate, the DM number density freezes out to the
current relic density of $\Omega h^{2} = 0.12$ \cite{RD}. The DM coannihilation
rate for $\phi _{1}\phi _{2} \to A^{\prime *} \to f\bar{f}$ is
\cite{Excited_DM}
%
%e3 #&#
\begin{align}
\label{equ:coannil}
\sigma \upsilon =
\frac{8\pi \epsilon ^{2}\alpha \alpha _{D} m_{1} m_{2} \upsilon ^{2}}{3\left [ \left ( m_{1}+m_{2}\right )^{2} - m_{A'}^{2}\right ]^{2}}
,
\end{align}
where $\alpha = e^{2}/4\pi $, $\alpha _{D} = g_{D}^{2}/4\pi $, and
$\upsilon $ is the relative velocity between $\phi _{1}$ and
$\phi _{2}$. Contrary to the case of fermionic DM particles, this process
is $p$-wave suppressed. Therefore, the scalar DM model provides an extra
handle to relax the tension between a light DM and the CMB result in which
a $s$-wave DM coannihilation is assumed \cite{RD}. As illustrated in
\cite{Excited_DM,monophoton_plus_ee}, the existing experimental limits
favor a small mass splitting between $\phi _{1}$ and $\phi _{2}$, and to
avoid DM overabundance, the masses of $\phi _{1,2}$ close to the threshold,
$m_{A'} \simeq m_{1} + m_{2}$, are preferred. This has two consequences:
$\phi _{1}$ produced at colliders tends to have a longer lifetime, and
its flight direction is very much collinear with its mother dark photon.
Because an on-shell dark photon is mainly produced in the forward detector
direction (see Sec.~\ref{sec:prod}), we propose putting a DM detector in
the forward region to detect the decay products of $\phi _{1}$, while for
an off-shell dark photon, we propose putting it in the central region.
To check the relic density compatibility, we employ \text{\texttt{micrOMEGAs}}
\cite{micrOMEGAs} to perform numerical calculations of the density evolution.

%s4 #&#
\section{Dark photon production at $e^{+}e^{-}$ colliders}
\label{sec:prod}

\begin{figure}
\centering
\includegraphics[width=0.235\textwidth]{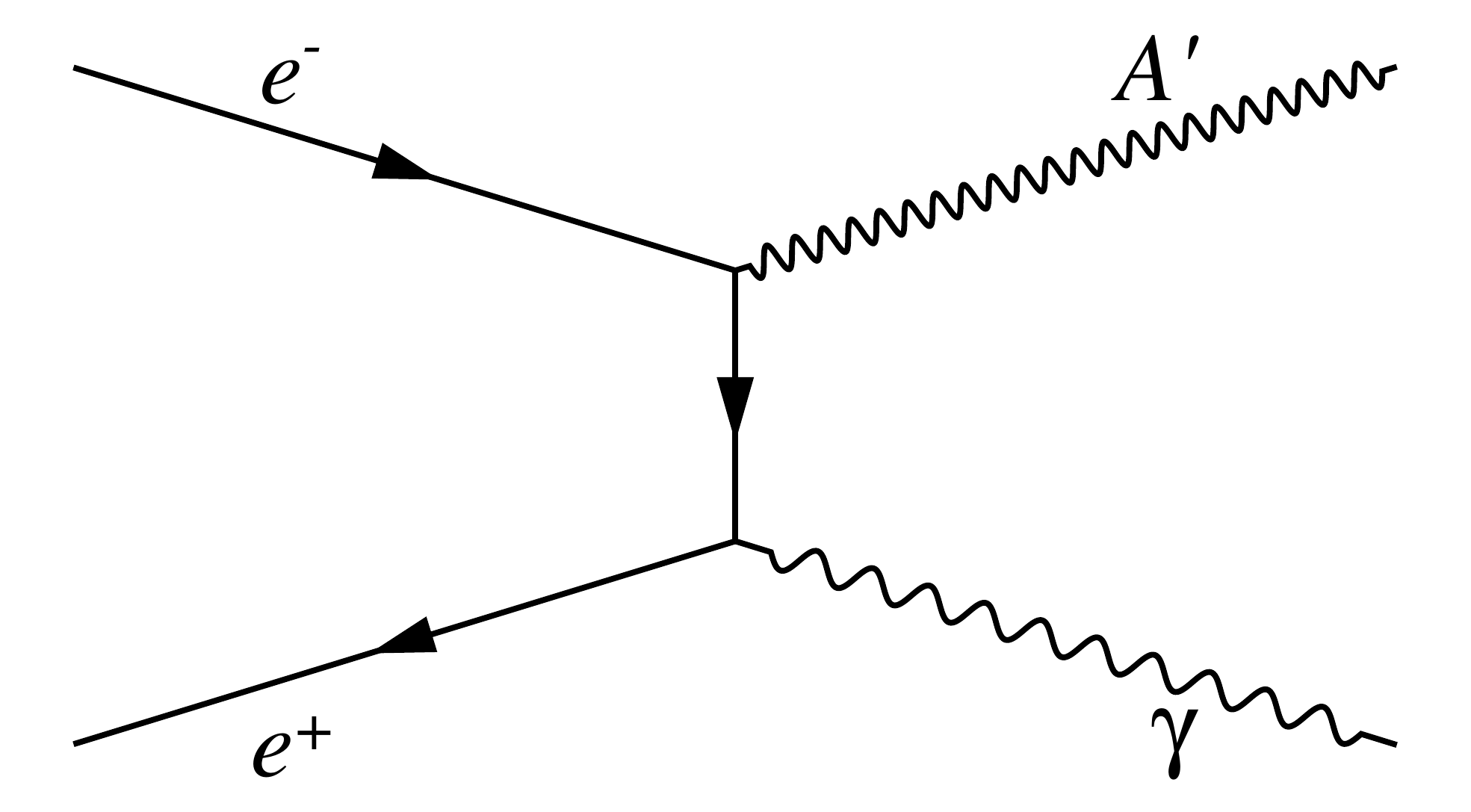}
\includegraphics[width=0.235\textwidth]{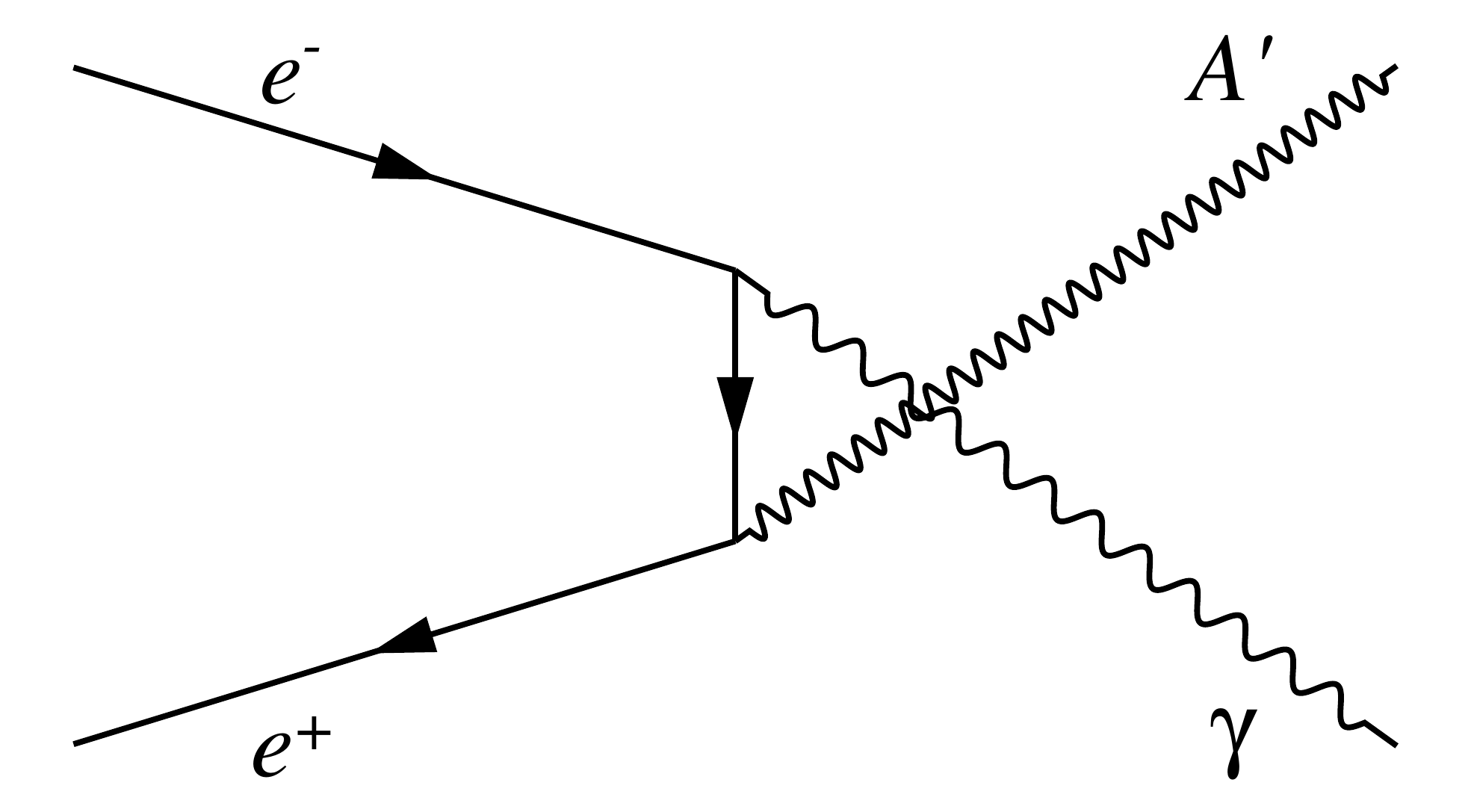}
\caption{The production diagrams of an on-shell dark photon $A'$ at a
$e^{+}e^{-}$ collider.}
\label{fig:prod}
\end{figure}

The main production diagrams of a dark photon together with an Initial
State Radiation (ISR) photon, i.e., $e^{+}e^{-}\to \gamma A'$, at a
$e^{+}e^{-}$ collider are shown in Fig.~\ref{fig:prod}.

%f1 #&#

The Born level differential cross section of the process
$e^{+}e^{-}\to A'\gamma $ with respect to $\cos \theta $, where
$\theta $ is the polar angle between the outgoing $A'$ and the electron
beam axis, can be expressed as \cite{QiaoCongfeng}
%
%e4 #&#
\begin{align}
\label{equ:prod_X}
\frac{d\sigma }{d\cos \theta } = 2\pi \epsilon ^{2}\alpha ^{2}
\frac{\left (s+m_{A'}^{2}\right )^{2}+\left (s-m_{A'}^{2}\right )^{2} \cos ^{2}\theta }{s\left (s-m_{A'}^{2}\right )\left (s \sin ^{2}\theta + 4m_{e}^{2} \right )},
\end{align}
where $s$ is the center of mass (CM) energy squared of the
$e^{+}e^{-}$ system, and $m_{e}$ the electron mass. According to this expression,
it can be seen that the dark photon is primarily produced in the very forward
directions of a $e^{+}e^{-}$ detector system. The production cross section
increases with dark photon mass for a given $\sqrt{s}$, while for a given
dark photon mass, it decreases with $\sqrt{s}$.

\begin{figure}
\centering
\includegraphics[width=0.30\textwidth]{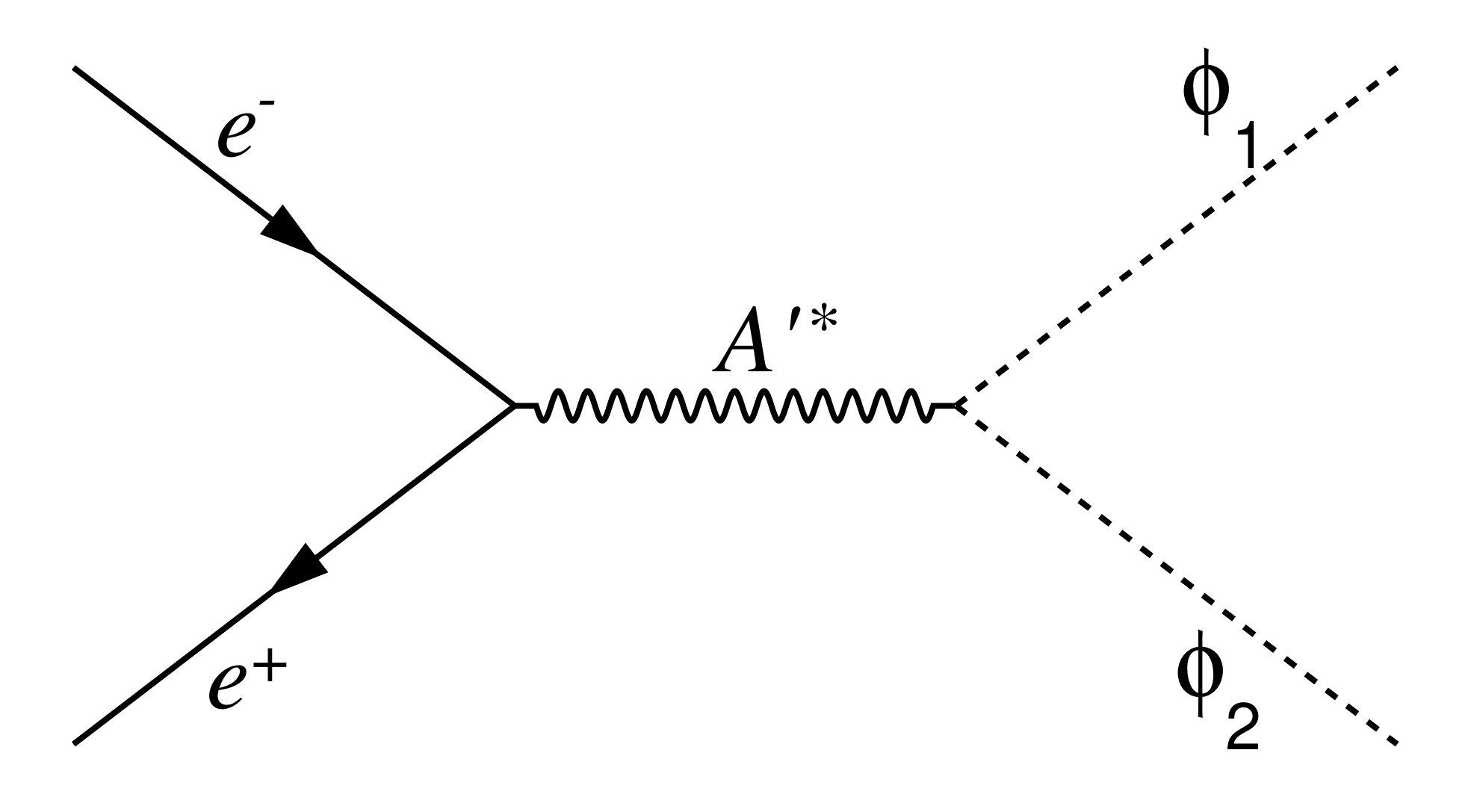}
\caption{The production diagrams of an off-shell dark photon $A'$ at a
$e^{+}e^{-}$ collider.}
\label{fig:prod_off}
\end{figure}

Apart from the on-shell $A'$ production with an ISR photon, the DM particles
can be directly produced through an off-shell $A'$, as shown in Fig.~\ref{fig:prod_off}. The differential cross section reads
%
%e5 #&#
\begin{align}
\label{equ:prod_X_off}
\frac{d\sigma }{d\cos \theta } = &
\frac{\epsilon ^{2} \alpha g_{D}^{2} s}{32 \left (s-m_{A'}^{2}\right )^{2}}
\left ( 1-\frac{(m_{1}+m_{2})^{2}}{s} \right )^{\frac{3}{2}}
\nonumber
\\
& \cdot \left ( 1-\frac{(m_{1}-m_{2})^{2}}{s} \right )^{\frac{3}{2}}
\sin ^{2}\theta .
\end{align}
The striking difference with respect to Eq.~\ref{equ:prod_X} is that
$\phi _{1}$ will be most abundantly produced in the central region of the
detector.\footnote{In the case of fermionic DM, for $m_{1}+m_{2}$ close to
$\sqrt{s}$ favored by the relic density, the angular probability of
$\phi _{1}$ is more isotropic.} If $\phi _{1}$ is long-lived, no trace
will be left inside the $e^{+}e^{-}$ detector, whereas if it is short-lived,
a pair of non-resonant soft tracks with a displaced vertex can be left
inside, but they are very hard to be detected due to the collision background.

%f2 #&#

%s5 #&#
\section{Dark photon decay}

After $A'$ is produced, it then promptly decays into a charged SM fermion
pair, or dark scalars $\phi _{1}$ and $\phi _{2}$. The spin averaged
$A'$ decay width to each final state can be expressed as
%
%e6 #&#
%e7 #&#
\begin{align}
\label{equ:DPh_wid}
\Gamma(A'\to f\bar{f})  = & \frac{1}{3} \epsilon^2 Q_f^2 \alpha m_{A'} \left(1+\frac{2m_f^2}{m_{A'}^2} \right)  \left(1-\frac{4m_f^2}{m_{A'}^2} \right)^{\frac{1}{2}}, \\
\Gamma(A'\to \phi_1\phi_2)  = & \frac{g_D^2}{48\pi}m_{A'}\left( 1 - \frac{2(m_1^2+m_2^2)}{m_{A'}^2} + \frac{m_1^4+m_2^4-2m_1^2m_2^2}{m_{A'}^4} \right) \nonumber \\
& \cdot \left( 1-\frac{(m_1+m_2)^2}{m_{A'}^2} \right)^{\frac{1}{2}} \left( 1-\frac{(m_1-m_2)^2}{m_{A'}^2} \right)^{\frac{1}{2}} ,
\end{align}
where $m_{f}$ denotes the charged fermion mass ($f=e,\mu ,u,d,s$). The partial decay width into
quarks can be collectively named $\Gamma _{had}$, and can be effectively
evaluated using the experimental $R$ data \cite{PDG}, which is defined
as
%
%e8 #&#
\begin{align}
\label{equ:R}
R(\sqrt{s}) = \frac{\sigma(e^+e^-\to\text{hadrons})}{\sigma(e^+e^-\to\mu^+\mu^-)}.
\end{align}
This experimental data input starts from $\sqrt{s}=0.3$ GeV (just above
the $\pi \pi $ mass threshold), with which $\Gamma _{had}$ can be expressed
as
%
%e9 #&#
\begin{align}
\label{equ:DPh_wid_had}
\Gamma _{had} = \Gamma _{\mu \mu } \cdot R(\sqrt{s}=m_{A'}).
\end{align}
Since a small mass splitting is preferred by the relic density constraint,
the signal mass range we consider would mostly mean a mass splitting below
the $\rho ^{0}$ meson threshold. This means that the hadronic contribution
will be small, and the hadronic final state will be dominated by a
$\pi ^{+}\pi ^{-}$ pair. When $g_{D}$ is orders of magnitude larger than
$\epsilon $, $A'$ will predominantly decay into the dark final state
$\phi _{1}\phi _{2}$. Namely, the branching ratio (BR) of
$A'\to \phi _{1}\phi _{2}$ will be large, which is the case we consider.

%s6 #&#
\section{Heavier DM particle decay}

\begin{figure}
\centering
\includegraphics[width=0.34\textwidth]{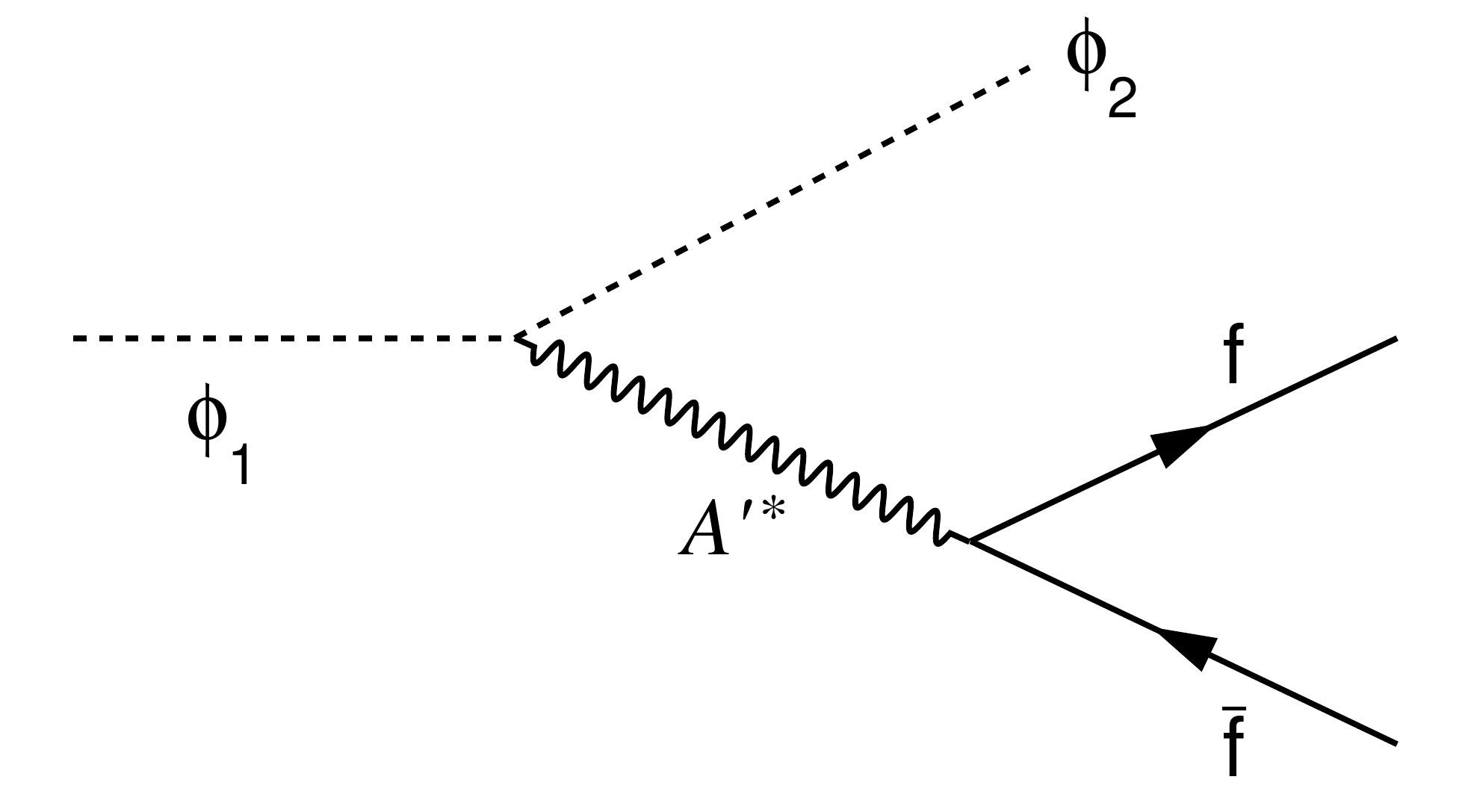}
\caption{The decay diagram of a heavier dark scalar $\phi _{1}$ into the lighter
$\phi _{2}$ and a $f\bar{f}$ pair through an off-shell $A^{\prime *}$.}
\label{fig:decay_1}
\end{figure}

The $\phi _{1}$ will fly for some distance, and then decays into
$\phi _{2}$ and a $f\bar{f}$ pair through an off-shell
$A^{\prime *}$, as shown in Fig.~\ref{fig:decay_1}.

%f3 #&#

\begin{figure}
\centering
\includegraphics[width=0.40\textwidth]{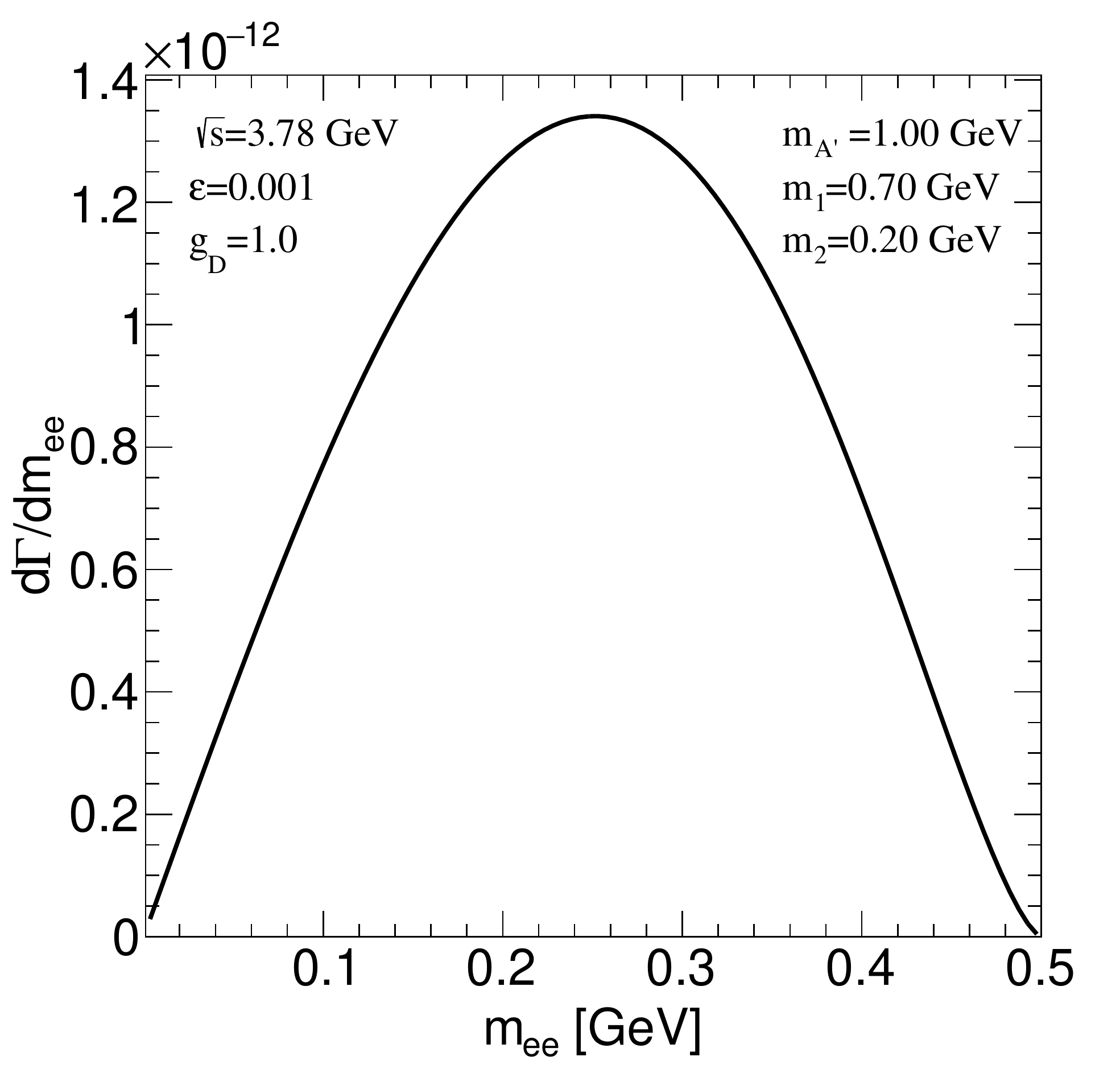}
\caption{The differential decay width of $\phi _{1}\to \phi _{2} e^{+}e^{-}$ as
a function of $m_{ee}$, with the chosen parameters indicated in the plot.}
\label{fig:decay_3body}
\end{figure}

The partial decay width of the three-body decay
$\phi _{1}\to \phi _{2} f\bar{f}$ can be expressed as
%
%e10 #&#
\begin{align}
\label{equ:phi1_wid}
&d\Gamma (\phi _{1}\to \phi _{2} f\bar{f})
\nonumber
\\
& =
\frac{\epsilon ^{2} Q_{f}^{2} \alpha g_{D}^{2}}{8\pi ^{2} m_{1}^{3} (m_{f\bar{f}}^{2}-m_{A'}^{2})^{2}}
\left [ (m_{\bar{f}\phi _{2}}^{2}-m_{f}^{2}) (m_{1}^{2}+m_{2}^{2} +
\right .
\nonumber
\\
&\quad  \left . m_{f}^{2}-m_{f\bar{f}}^{2}-m_{\bar{f}\phi _{2}}^{2}) - m_{1}
m_{2} \right ] dm_{f\bar{f}}^{2} dm_{\bar{f}\phi _{2}}^{2},
\end{align}
where $m_{f\bar{f}}$ denotes the $f\bar{f}$ pair invariant mass, and
$m_{\bar{f}\phi _{2}}$ the invariant mass of $\bar{f}$ and
$\phi _{2}$. After integrating out $m_{\bar{f}\phi _{2}}$, the partial
decay width becomes
%
%e11 #&#
\begin{align}
\label{equ:phi1_wid_2}
&\frac{d\Gamma _{\phi _{1}}}{dm_{f\bar{f}}}
\nonumber
\\
&=
\frac{\epsilon ^{2} Q_{f}^{2} \alpha g_{D}^{2}}{24\pi ^{2} m_{1}^{3} (m_{f\bar{f}}^{2}-m_{A'}^{2})^{2} m_{f\bar{f}}^{2}}
\left [ (m_{1}^{2}-m_{2}^{2}-m_{f\bar{f}}^{2})^{2} (m_{f\bar{f}}^{2}+2m_{f}^{2})
\right .
\nonumber
\\
&\quad  \left . -4m_{2}^{2} m_{f\bar{f}}^{4} - 8m_{2}^{2}m_{f}^{2}m_{f
\bar{f}}^{2} \right ] \left (m_{f\bar{f}}^{2}-4m_{f}^{2}\right )^{
\frac{1}{2}} \left [ (m_{1}^{2}-m_{2}^{2}
\right .
\nonumber
\\
&\quad  \left .
-m_{f\bar{f}}^{2})^{2}
-4m_{2}^{2} m_{f\bar{f}}^{2} \right ]^{\frac{1}{2}} ,
\end{align}
and the allowed range for $m_{f\bar{f}}$ is
$2m_{f}\leq m_{f\bar{f}}\leq (m_{1}-m_{2})$. When $f=e$, the differential
decay width of $\phi _{1}$ is shown in Fig.~\ref{fig:decay_3body}. It is
evident that the $ee$ pair mass has a broad spectrum spanning the whschemole
allowed mass range. For the hadronic three-body decays of
$\phi _{1}$, as for the case of $A'$, the $R$ input is used.
%
%f4 #&#

%f5 #&#
\begin{figure}
\centering
\includegraphics[width=0.45\textwidth]{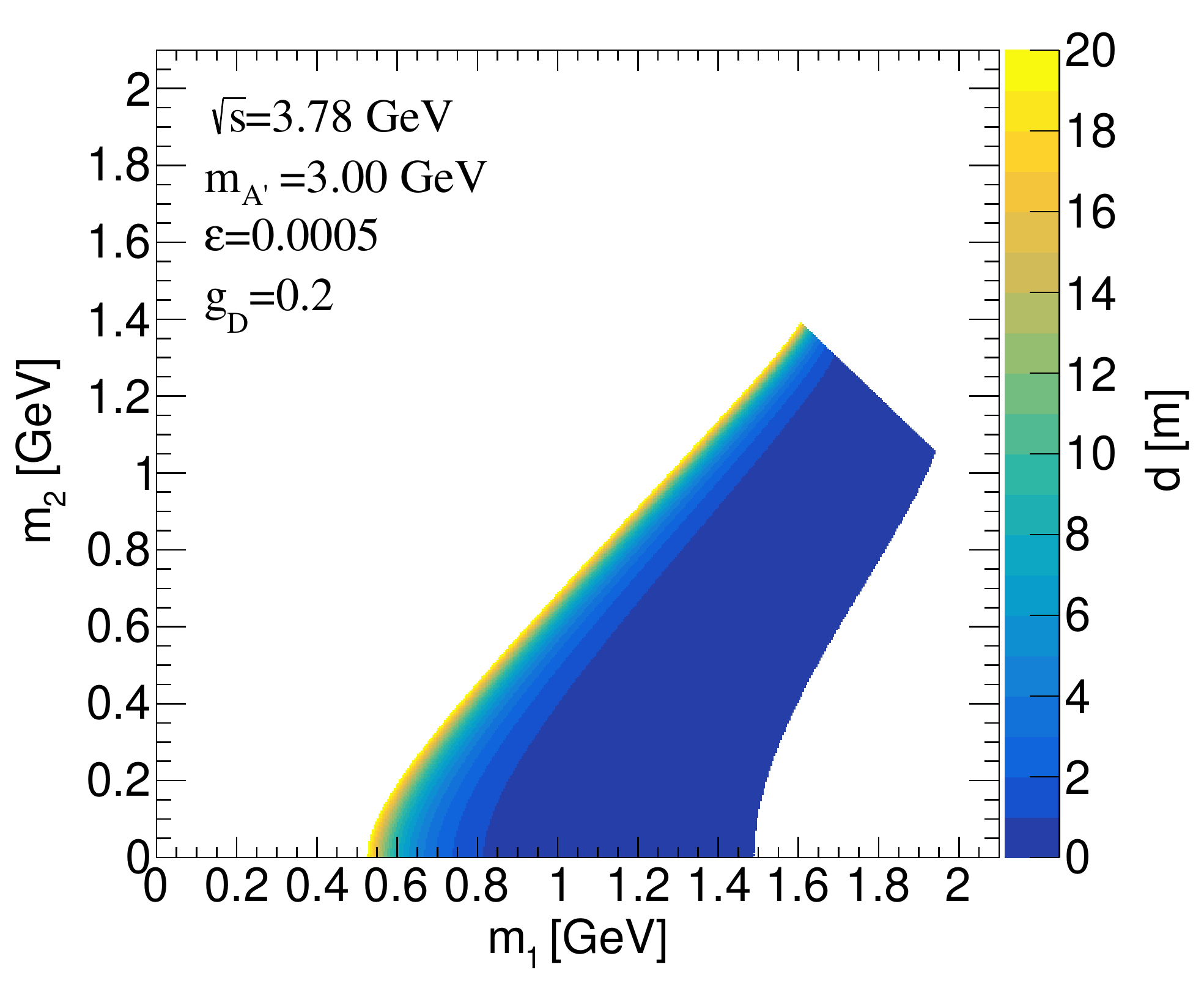}
\caption{The distribution of $\phi _{1}$ average flight length $d$ as a function
of $m_{1}$ vs. $m_{2}$, with the chosen parameter values indicated in the plot.
The $\phi _{1}$ Lorentz boost factor is based on
Eq.~\ref{equ:energy_phi1} but averaged over $\theta _{1}^{*}$.}
\label{fig:d_m1m2_1}
\end{figure}

Since it is produced from the $A'$ decay, the flight length ($d$) of
$\phi _{1}$ is determined by its Lorentz boost factor $\gamma _{1}$ and
its total decay width by
%
%e12 #&#
\begin{align}
\label{equ:phi1_d}
d =
\frac{\gamma _{1} \beta _{1}}{\sum _{f}\Gamma _{\phi _{1}\to \phi _{2} f\bar{f}}},
\end{align}
where $\beta _{1}$ is the speed of $\phi _{1}$. The $\phi _{1}$ boost factor
varies from event to event, but an average value can be obtained by averaging
over the angle $\theta _{1}^{*}$ in Eq.~\ref{equ:energy_phi1} introduced
in the next section. The $d$ distributions as a function of $m_{1}$ and
$m_{2}$ are shown in Fig.~\ref{fig:d_m1m2_1}, in the setting of BESIII. It is evident that as the
mass splitting $m_{1}-m_{2}$ increases, the flight length $d$ decreases,
and the region with $d<1$ m is best suitable for the displaced vertex search
in \cite{monophoton_plus_ee}. On the other hand, as mentioned in Sec.~\ref{sec:RD}, the DM relic density constraint favors a small mass splitting
(hence a larger $d$), which is the focus in this work.

%s7 #&#
\section{Heavier DM particle distributions in the on-shell case}

The angular distribution of $\phi _{1}$ from an off-shell $A'$ decay is
already given in Eq.~\ref{equ:prod_X_off}. For the case of on-shell
$A'$, the double differential cross section for the $\phi _{1}$ production,
in the narrow width approximation and with full spin correlation between
$e^{+}e^{-}\to \gamma A'$ and $A'\to \phi _{1}\phi _{2}$, is proportional
to
%
%e13 #&#
\begin{align}
\label{equ:prod_phi1}
\frac{d\sigma }{d\Omega d\Omega _{1}^{*}} \propto &
\frac{1}{s\left ( s\sin ^{2}\theta + 4 m_{e}^{2}\right )} \left \{
\left (s+m_{A'}^{2}\right )^{2} + \left (s-m_{A'}^{2}\right )^{2}
\cos ^{2}\theta \right .
\nonumber
\\
& - \left [ \left (s-m_{A'}^{2}\right )^{2} + \left (s+m_{A'}^{2}
\right )^{2}\cos ^{2}\theta \right ] \cos ^{2}\theta _{1}^{*}
\nonumber
\\
& + 4m_{A'} \sqrt{s} \left (s+m_{A'}^{2}\right ) \sin \theta _{1}^{*}
\cos \theta _{1}^{*} \cos \phi _{1}^{*} \sin \theta \cos \theta
\nonumber
\\
& \left . - 4m_{A'}^{2} s \sin ^{2}\theta _{1}^{*} \cos ^{2}\phi _{1}^{*}
\sin ^{2}\theta \right \}  ,
\end{align}
where $\Omega $ ($\Omega _{1}^{*}$) is the solid angle of $A'$ ($
\phi _{1}$) in the $e^{+}e^{-}$ lab ($A'$ rest) frame, and
$\theta _{1}^{*}$ and $\phi _{1}^{*}$\footnote{Note that
$\phi _{1}^{*}$ is also the angle between the plane formed by the beam
axis and outgoing $A'$, and the plane formed by the $A'$ decay products.}
are the polar and azimuth angles of $\Omega _{1}^{*}$, with the flight
direction of $A'$ in the $e^{+}e^{-}$ lab frame as the polar axis for
$\Omega _{1}^{*}$.

Contrary to Eq.~\ref{equ:prod_X} where $A'$ is produced primarily in the
forward ($\theta =0$ or $\pi $) region, it is seen from Eq.~\ref{equ:prod_phi1} that when $\theta \to 0$ or $\pi $, and
$\theta _{1}^{*}\to 0$, the cross section vanishes, which creates a ``hole''
in the very forward region. With $\Omega $ and $\Omega _{1}^{*}$ specified,
the energy of $\phi _{1}$ in the $e^{+}e^{-}$ lab frame can be expressed
as
%
%e14 #&#
\begin{align}
\label{equ:energy_phi1}
E_{1} = & \frac{s+m_{A'}^{2}}{4\sqrt{s}} \left (1+
\frac{m_{1}^{2} - m_{2}^{2}}{m_{A'}^{2}}\right ) +
\nonumber
\\
& \frac{s-m_{A'}^{2}}{4\sqrt{s}} \left [ \left (1+
\frac{m_{1}^{2} - m_{2}^{2}}{m_{A'}^{2}}\right )^{2} -
\frac{4m_{1}^{2}}{m_{A'}^{2}} \right ]^{\frac{1}{2}} \cos \theta _{1}^{*},
\end{align}
and the angle between its flight direction and the electron beam axis in
the same frame ($\theta _{1}$) can be determined via
%
%e15 #&#
\begin{align}
\label{equ:ang_phi1}
\cos \theta _{1} = & \left \{  \left [s-m_{A'}^{2}+(s+m_{A'}^{2})
\beta _{1}^{*}\cos \theta _{1}^{*}\right ]\cos \theta - \right .
\nonumber
\\
& \left . 2\sqrt{s} m_{A'}\beta _{1}^{*}\sin \theta _{1}^{*}\cos
\phi _{1}^{*}\sin \theta \right \}  \cdot \left \{  4s m_{A'}^{2}
\beta _{1}^{*2}\sin ^{2}\theta _{1}^{*} + \right .
\nonumber
\\
& \left . \left [s-m_{A'}^{2}+(s+m_{A'}^{2})\beta _{1}^{*}\cos
\theta _{1}^{*}\right ]^{2} \right \}  ^{-\frac{1}{2}},
\end{align}
where $\beta _{1}^{*}$ is a constant defined as
%
%e16 #&#
\begin{align}
\label{equ:beta_1_s}
\beta _{1}^{*} = \left ( 1-
\frac{4m_{A'}^{2} m_{1}^{2}}{\left (m_{A'}^{2}+m_{1}^{2}-m_{2}^{2} \right )^{2}}
\right )^{\frac{1}{2}}.
\end{align}

\begin{figure}
\centering
\includegraphics[width=0.45\textwidth]{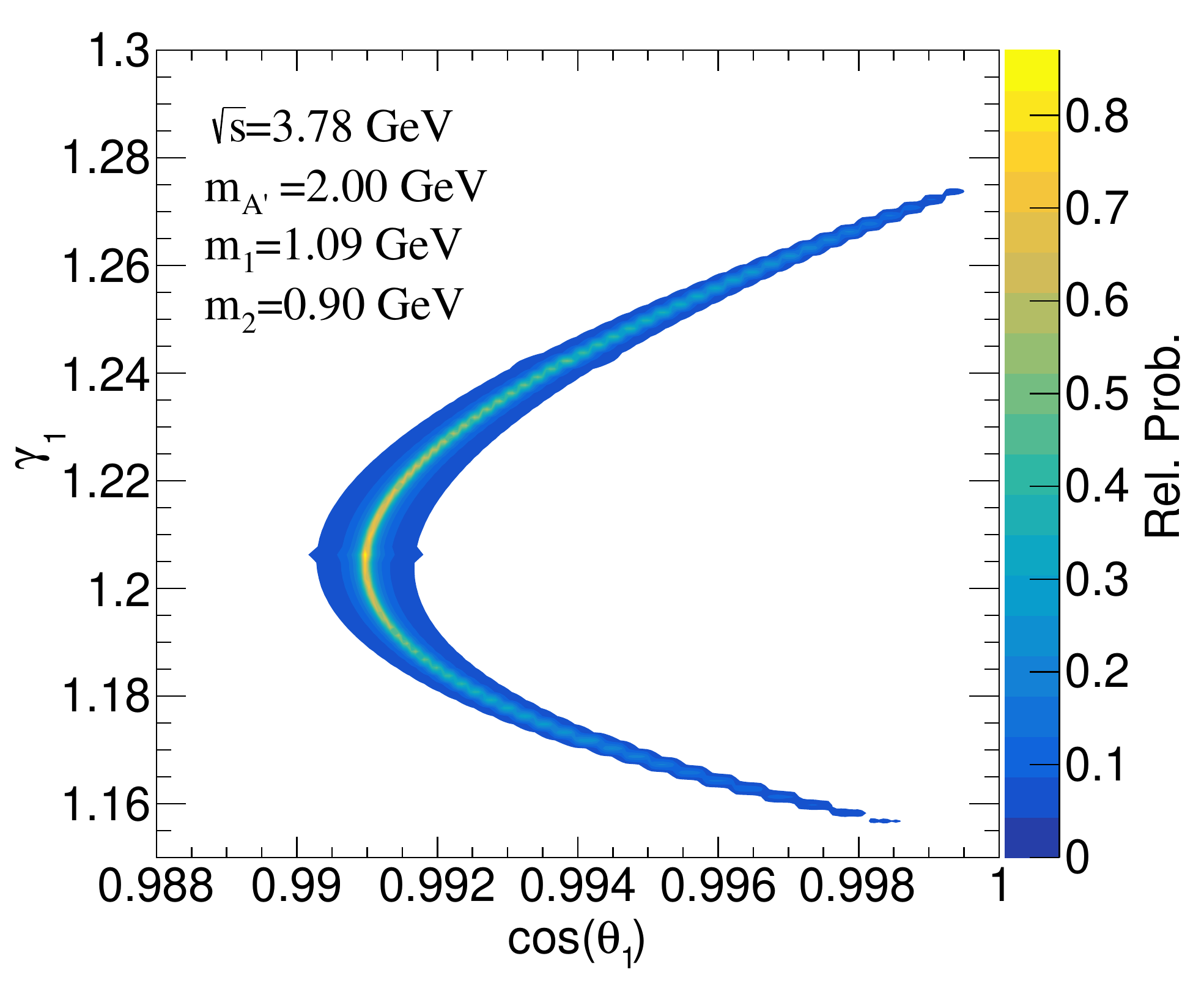}
\caption{The 2-D probability distribution of $\cos \theta _{1}$ vs. Lorentz
boost factor $\gamma _{1}$ of $\phi _{1}$ in the $e^{+}e^{-}$ lab frame, with
the chosen parameters indicated in the plot. The color scale indicates relative
probability.}
\label{fig:2D_phi1_par1}
\end{figure}

\begin{figure}
\centering
\includegraphics[width=0.40\textwidth]{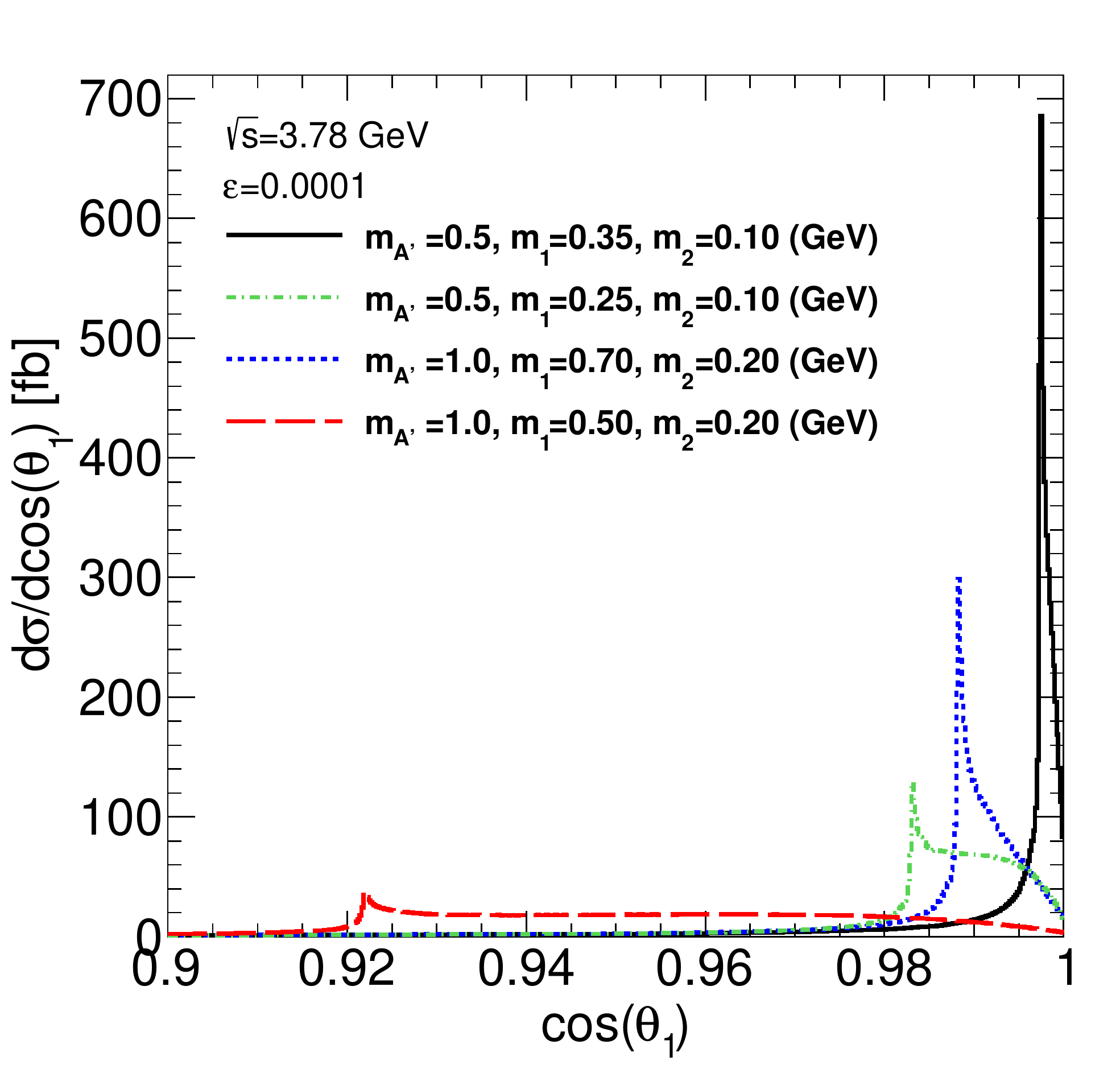}
\caption{The differential cross sections of $\phi _{1}$ as a function of $\cos
\theta _{1}$ in the $e^{+}e^{-}$ lab frame, corresponding to four different sets
of chosen parameters indicated in the plot.}
\label{fig:cosTheta_phi1}
\end{figure}

Based on Eq.~\ref{equ:prod_phi1}-\ref{equ:beta_1_s}, the 2-D probability
density distribution of $\phi _{1}$ Lorentz boost factor
$\gamma _{1}$ and its $\cos (\theta _{1})$ in the lab frame for a particular
parameter set is shown in Fig.~\ref{fig:2D_phi1_par1},\footnote{The irregular
boundaries are numerical artifacts.} and the differential cross section
of $\phi _{1}$ as a function of $\cos (\theta _{1})$ is shown in Fig.~\ref{fig:cosTheta_phi1} in the vicinity of $\theta _{1}=0$.

%f6 #&#

%f7 #&#

%s8 #&#
\section{Dark matter decay detector}
\label{sec:det}

\begin{figure*}
\centering
\includegraphics[width=0.35\textwidth]{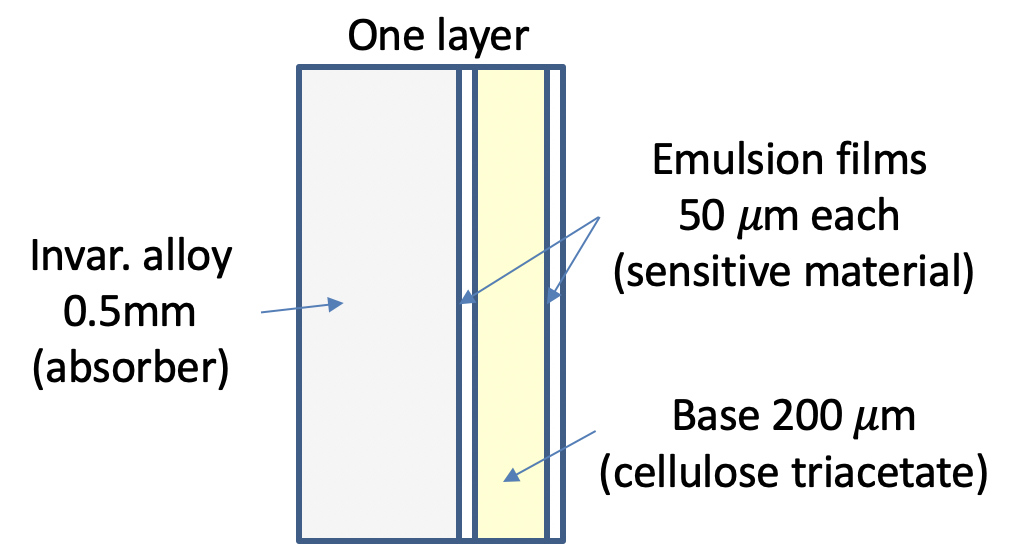} \hspace{10mm}
\includegraphics[width=0.35\textwidth]{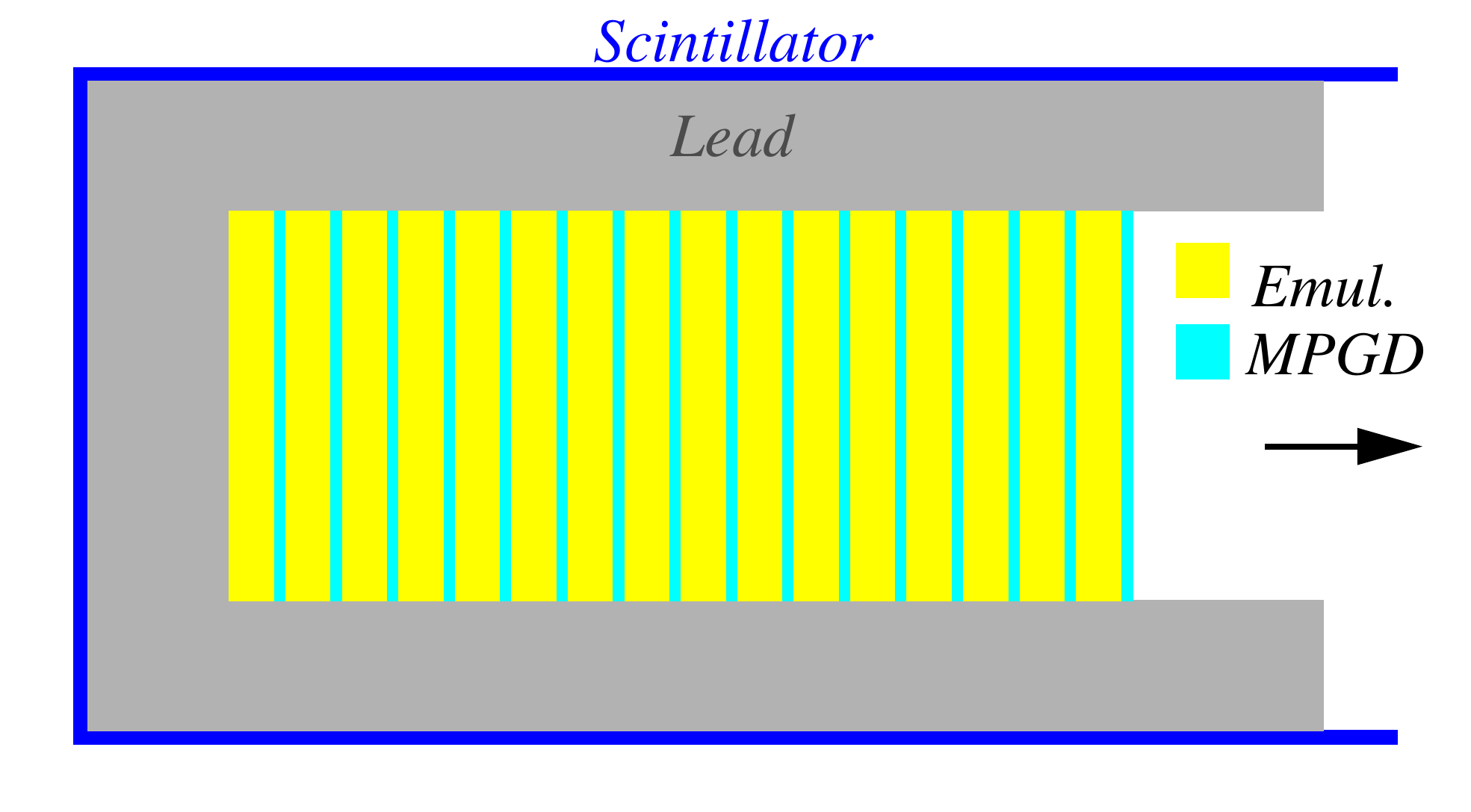}
\put(-235, 84){\textbf{(a)}}
\put(-16, 80){\textbf{(b)}}
\caption{The illustration of an emulsion layer (a) and the cross sectional
sketch of the proposed detector (b).}
\label{fig:emulsion_MPGD}
\end{figure*}

To detect the visible decay products of $\phi _{1}$, we propose to use
an emulsion plus Micro-Pattern Gaseous Detector (MPGD) design. Emulsion
detectors have been used in OPERA \cite{Opera}, DsTau \cite{DsTau} and
FASER$\nu $ \cite{FASERnu} experiments. Our emulsion design has a multitude
of thin layers, each of which consists of a 0.5 mm thick invariant alloy
(consists of 63.8\% iron, 36.0\% nickel and 0.2\% carbon) plate and 0.3
mm thick emulsion sheet. The invariant alloy has the advantage of a low
thermal expansion coefficient, which makes the emulsion sheets alignment
less susceptible to temperature variations. Each emulsion sheet consists
of a 200~$\mu$m thick base (made of, e.g., cellulose acetate), sandwiched
between two emulsion films of 50~$\mu$m thickness, as illustrated in Fig.~\ref{fig:emulsion_MPGD}(a). The emulsion consists mainly of AgBr and gelatin
material. The silver bromide crystals are sensitive to ionization by charged
particles passing through the emulsion, and have a typical size of 0.2~$\mu$m. Therefore, the emulsion can have a position measurement of tracks
with a precision below 1~$\mu$m, which makes it ideal for our purpose
of detecting a pair of charged particles ($ee$, $\mu \mu $ or
$\pi \pi $) with a common vertex.

The AgBr crystal has a band gap of about 2.5 eV. Photons radiated from
charged particles incident on the AgBr crystal grain produce electrons
which interact to yield silver atoms. More photons will induce a larger
concentration of silver atoms, containing from 4 to tens of silver atoms,
which form an invisible latent image. During film development, the latent
image is amplified chemically by an order of $\sim $10$^{8}$ and silver
filaments are formed. After image fixing (removing the remaining silver
salts), these silver filaments can be read out under microscopes. At the
reconstruction level, sequences of aligned grains will be recognized and
form tracks of charged particles. With the Geant4 \cite{Geant4} simulation it is found
that, on average, a 10 MeV electron or positron travels through about 5
emulsion sheets, which amounts to 10 mini-track measurements (each emulsion
sheet has two films on each side). This can be a threshold cut on the tracks
in this work.

%f8 #&#

%f9 #&#
\begin{figure*}
\centering
\includegraphics[width=0.55\textwidth]{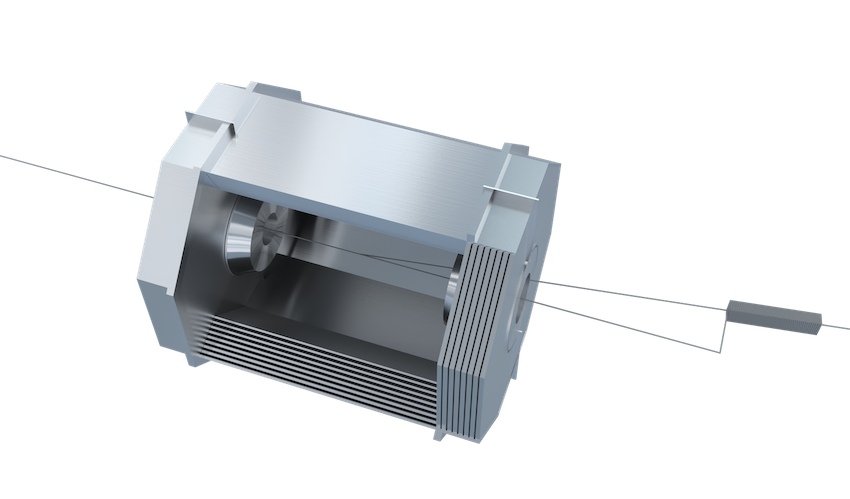}
\includegraphics[width=0.35\textwidth]{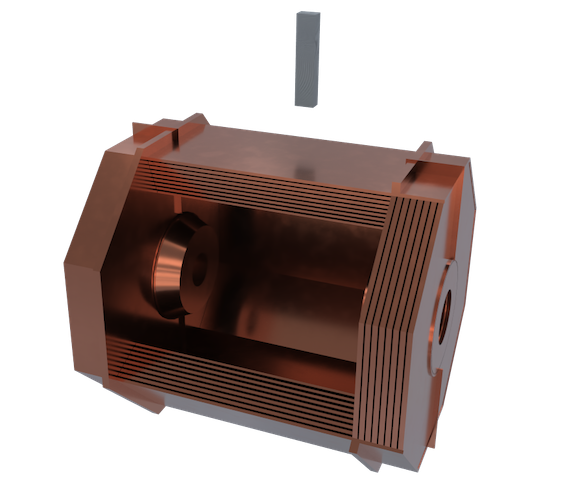}
\put(-355, 150){\textbf{(a)}}
\put(-135, 150){\textbf{(b)}}
\caption{(a) The schematic view of the DM detector's position (6 m away from the
IP, and 0.84 m off from the beam axis) relative to the BESIII detector (only the
muon identifier is shown). (b) The position of a DM detector placed in the
central region (4 m away from the IP) relative to the BESIII or Belle II
detector.}
\label{fig:cad}
\end{figure*}

%f10 #&#
\begin{figure*}
\centering
\includegraphics[width=0.41\textwidth]{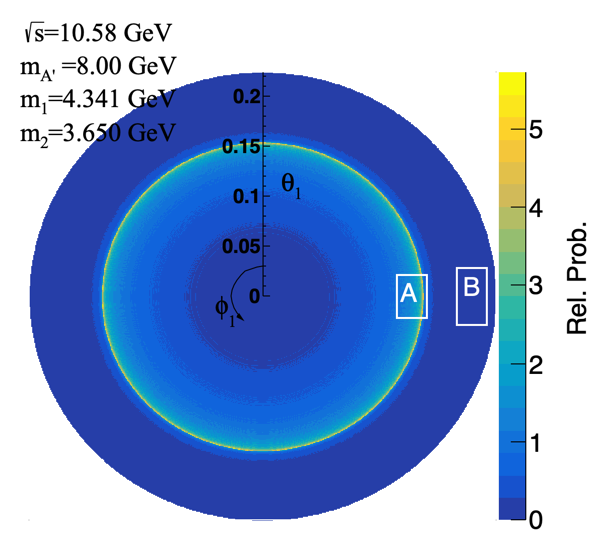} \hspace{2em}
\includegraphics[width=0.41\textwidth]{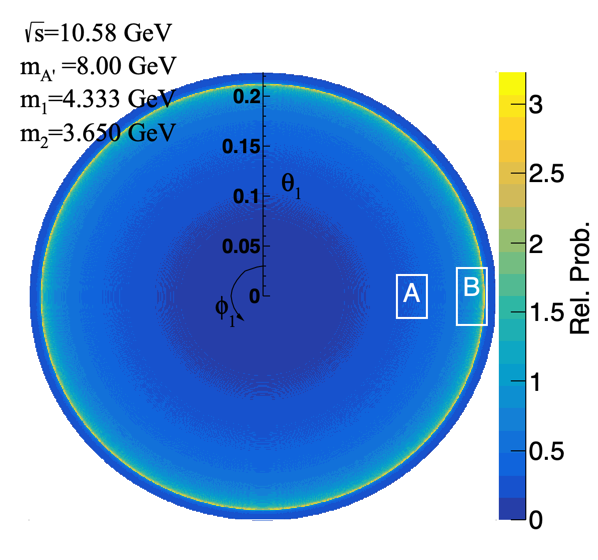}\\
\caption{The probability distribution of polar vs. azimuthal angle of particle
$\phi _{1}$ in the $e^{+}e^{-}$ lab frame, with the chosen parameters for an
on-shell $A'$ indicated in the plot. The cross sectional areas of two DM
detectors are indicated by the white box $A$ (0.2$\times $0.3 m$^{2}$) and $B$
(0.2$\times $0.4 m$^{2}$). Detector $A$ ($B$) is placed 1 m (1.4 m) off the beam
axis. Both are 7 m away from the IP. The color scale indicates relative
probability.}
\label{fig:prob_theta1_phi1}
\end{figure*}

%f11 #&#
\begin{figure*}[!htb]
\centering
\includegraphics[width=0.201\textwidth]{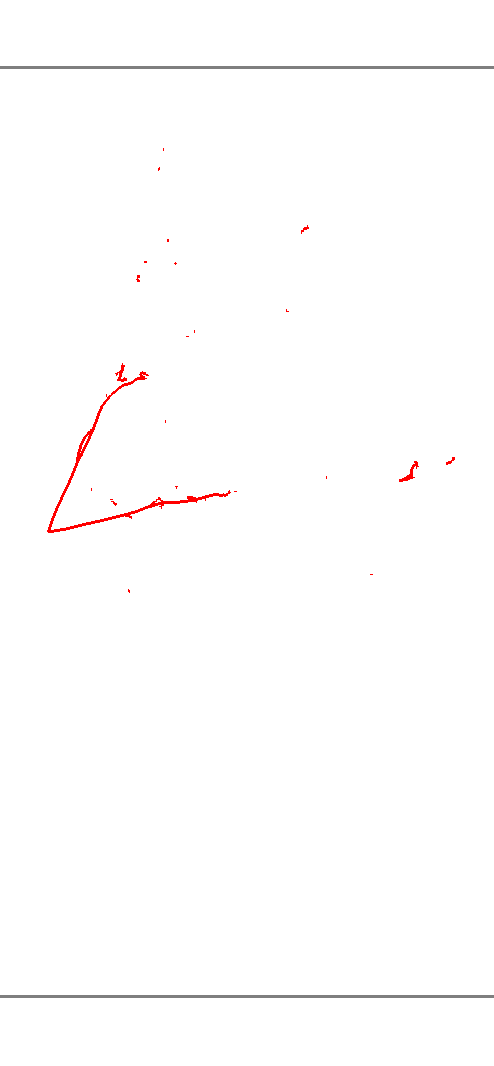}\hspace{2mm}
\includegraphics[width=0.21\textwidth]{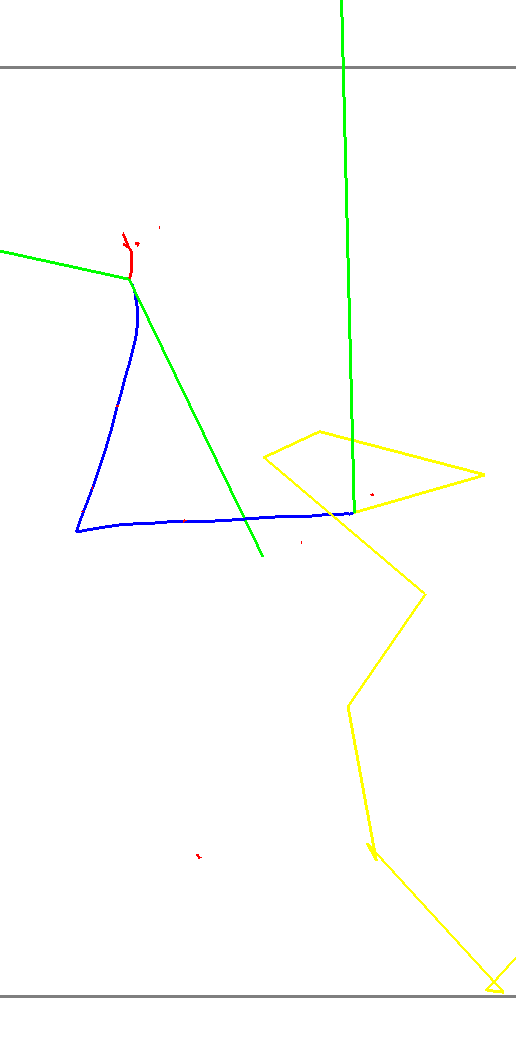}\hspace{2mm}
\includegraphics[width=0.2427\textwidth]{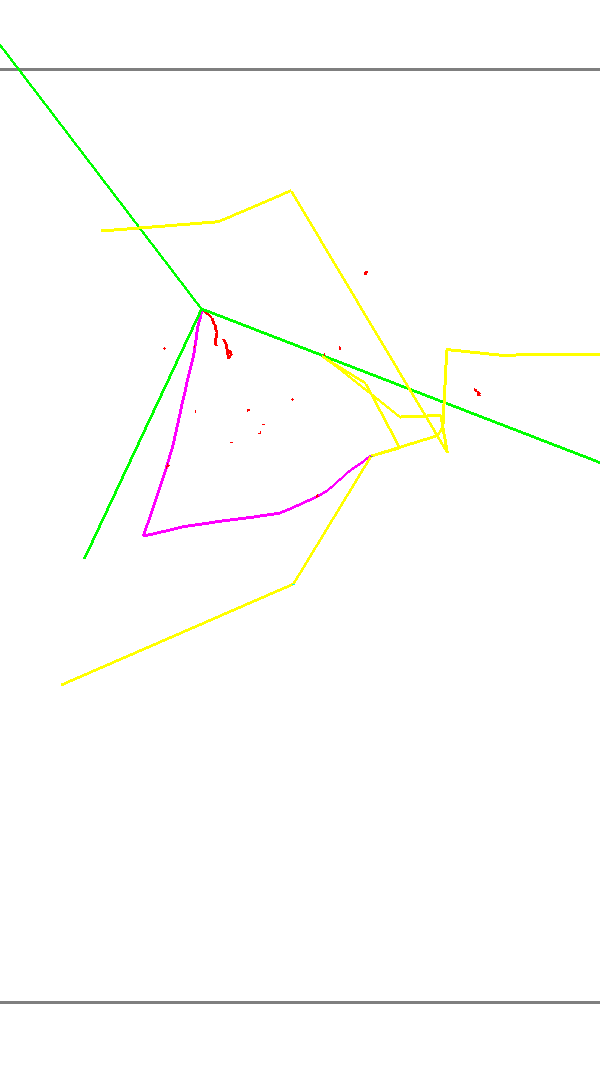}\hspace{2mm}
\includegraphics[width=0.176\textwidth]{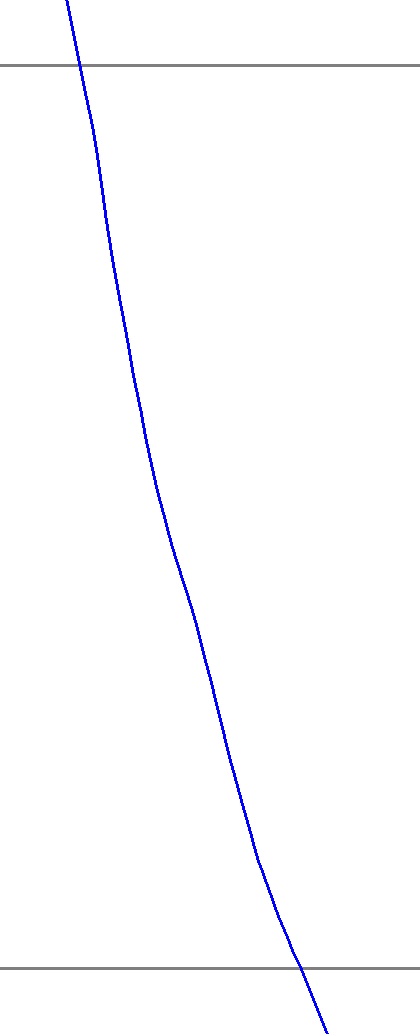}
\put(-410, 4.2){\textbf{(a)}}
\put(-295, 4.2){\textbf{(b)}}
\put(-170, 4.2){\textbf{(c)}}
\put(-50, 4.2){\textbf{(d)}}
\caption{The tracks of a $e^{+}e^{-}$ (a), $\mu ^{+}\mu ^{-}$ (b), $\pi ^{+}\pi
^{-}$ (c) pair coming from a $\phi _{1}$ decay, and a cosmic muon (d) in the DM
detector simulated with Geant4 (MPGD not included). The cosmic muon has an energy of 1 GeV, and all other tracks have a
kinetic energy of 100 MeV. The yellow
(green) trajectories represent neutrons (neutrinos), which leave little or no
energy in the detector. The cross sectional view along the $z$-axis is shown and
the horizontal gray lines indicate the top and bottom boundaries of the
detector.}
\label{fig:geant4}
\end{figure*}

%f12 #&#
\begin{figure}
\centering
\includegraphics[width=0.40\textwidth]{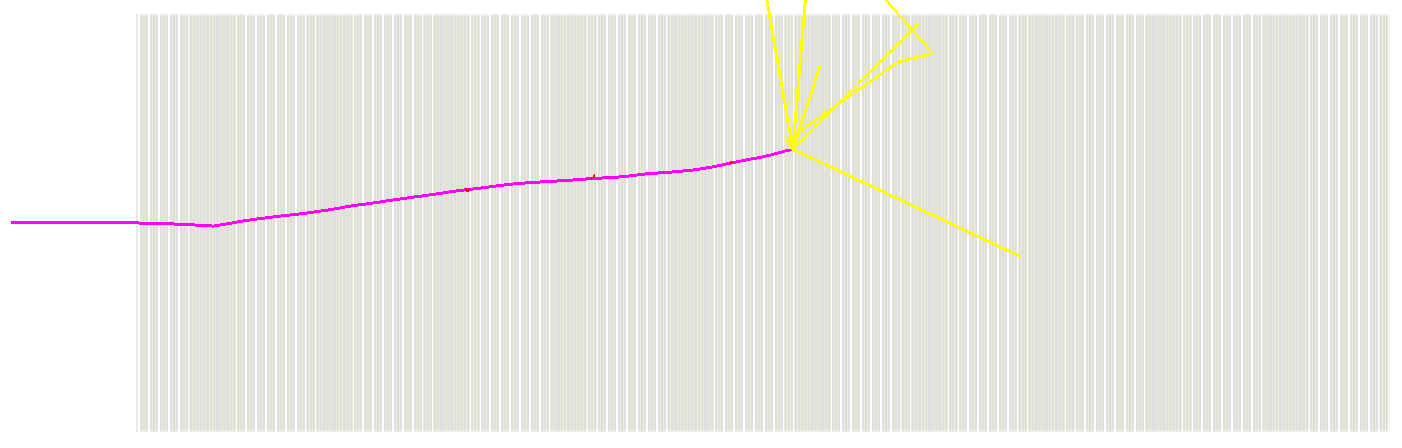}\\
\includegraphics[width=0.40\textwidth]{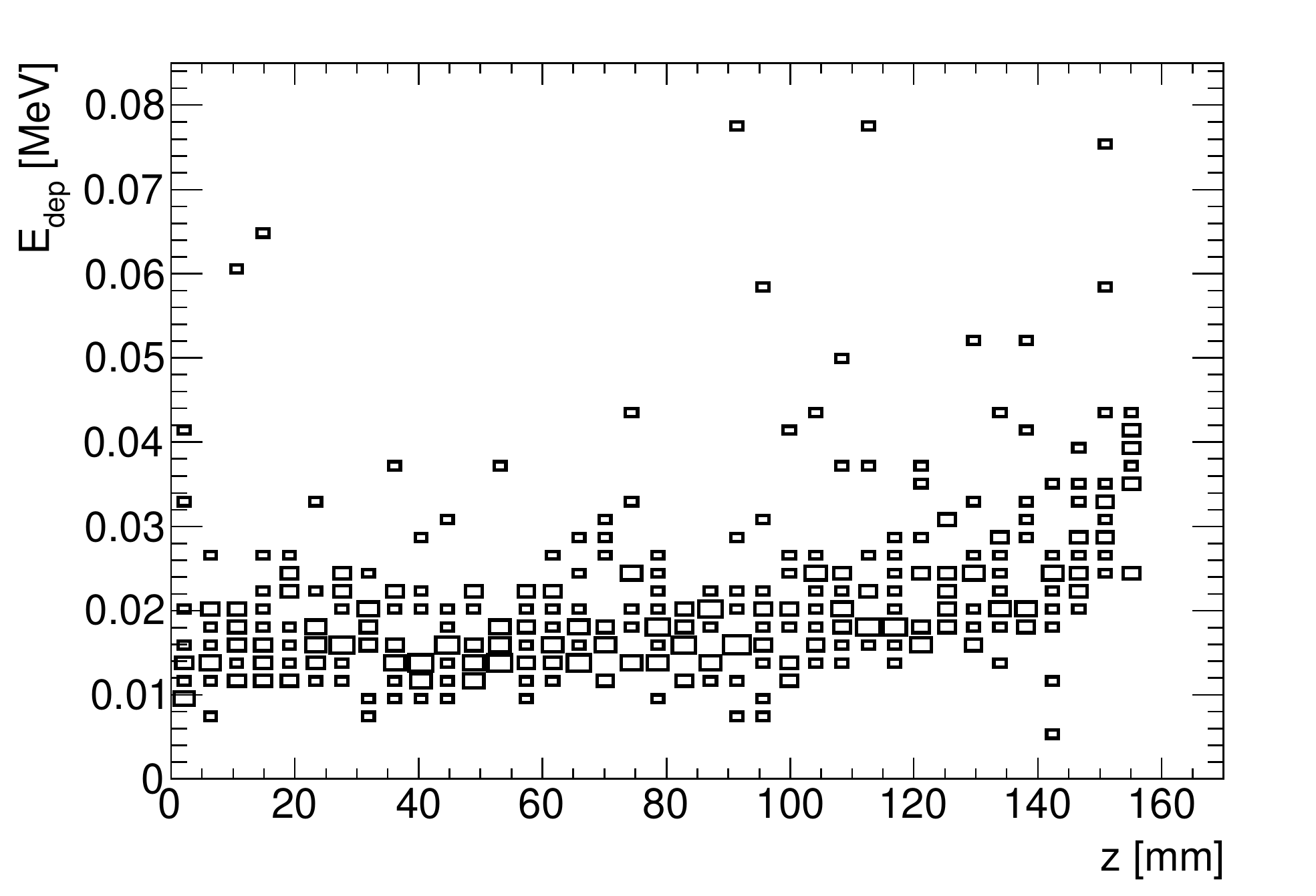}
\caption{The trajectory (in purple) of a 200 MeV (kinetic energy) charged pion
entering the detector along the $z$ direction from outside (top), and its energy
deposited in each emulsion film (bottom), based on the Geant4 simulation.}
\label{fig:geant4_pion}
\end{figure}

To distinguish backgrounds from cosmic muons and the associated secondary
particles (mostly photons, and a small amount of hadrons) produced by the
interaction between the muons and upstream material, thin MPGD layers are
sandwiched between blocks of emulsion layers. A suitable design of MPGD
that meet our requirement of thin thickness ($\lesssim $1 cm), good spatial
($\sim $50~$mu$m) and time resolution (a few ns) can be the Micro-Resistive
WELL ($\mu $RWELL) detector \cite{mRWELL}. Track segments in the MPGD layers
can be matched with those in the emulsion blocks to provide timing information.
If the time of a MPGD segment coincides with that of a cosmic muon or a
beam background photon, the matched track in the emulsion is not considered.
The time of cosmic muons can be obtained through scintillator fibers encasing
the whole active detector and the lead shield, as illustrated in Fig.~\ref{fig:emulsion_MPGD}(b). The time resolution of a scintillator is usually
at $\sim $1 ns level. Each emulsion block (yellow) contains 40 emulsion
layers and thus is 3.2 cm thick, while each MPGD layer (cyan) is 0.8 cm
thick. There is a 10 cm thick lead shield between the detector and outside
scintillator covering. This shield can prevent inner signal track from
reaching the scintillator, and can also help blocking photons from outside
sources (beam background) entering the detector, if it is put in a place
close to the beam. There can be 33 emulsion blocks and MPGD layers, making
the active part of the detector 1.32 m long. The cross sectional area of
the emulsion and MPGD can be 20 cm$\times $30 cm.

We propose to place the detector near a $e^{+}e^{-}$ collider such as BESIII
and Belle II. One possible placement around BESIII is illustrated in Fig.~\ref{fig:cad}(a) in the forward direction. The DM detector is facing toward
the interaction point (IP). For BESIII (Belle II), the distance between
the IP and its front area could be 6 m (7 m), and the perpendicular distance
between the front area center and the beam axis could be 0.84 m (1 m).\footnote{There
is a small beam crossing angle at the beam interaction region, and Belle
II beams are asymmetric in energy. These effects can be simulated, but
are not taken into account in this work for simplicity.} For the case
of an off-shell $A'$ and scalar DM, another placement is illustrated in
Fig.~\ref{fig:cad}(b), where the detector is placed 4 m away from the IP
in the central region facing the barrel. The probability distributions
of polar vs. azimuthal angle of particle $\phi _{1}$ in the
$e^{+}e^{-}$ lab frame in the forward direction are shown in Fig.~\ref{fig:prob_theta1_phi1}, where the areas covered by the proposed detector
are indicated by white rectangular boxes. The signal acceptance efficiencies
of the detector for some different parameters and detector positions are
given in Table~\ref{tab:acceptance}. Depending on specific model parameters,
acceptances in the two forward detectors can be very different. This enables
us to treat one as the signal detector, and the other as the background
control detector. Note that the same purpose can be fulfilled by putting
both forward detectors the same perpendicular distance off from the beam,
but at different lengths away from the IP. Although the central detector's
acceptance is much smaller than in the forward case (Table~\ref{tab:acceptance}), it can cover a signature that is very difficult
to be detected by the main $e^{+}e^{-}$ detector.

%t1 #&#
\begin{table}
\centering
\caption{ For the on-shell $A'$, the efficiencies of signal acceptance by the two DM detectors with two different mass parameter settings
, with $\sqrt{s}=10.58$ GeV and the detectors in the forward position as illustrated in Fig. \ref{fig:prob_theta1_phi1}. For the off-shel
l $A'$, the same efficiency with the DM detector placed in the central barrel position, as illustrated in Fig. \ref{fig:cad}(b). }
\label{tab:acceptance}
\begin{tabular}{c|cc}
\hline\hline
 On-shell case (GeV) & $\epsilon_A$ & $\epsilon_B$ \\ \hline
 $m_{A'}=8.0$, $m_1=4.341$, $m_2=3.650$ & $0.976\%$ & $0.065\%$ \\ \hline
 $m_{A'}=8.0$, $m_1=4.333$, $m_2=3.650$ & $0.194\%$ & $0.816\%$ \\ \hline\hline
 Off-shell case (GeV)  & \multicolumn{2}{c}{ $\epsilon_{off}$ } \\ \hline
 $m_{A'}=14.0$, $m_1=5.00$, $m_2=4.85$ & \multicolumn{2}{c}{ $0.045\%$ } \\ \hline
\hline
\end{tabular}
\end{table}

The tracks from $\phi _{1}$ decay inside the detector are simulated by
Geant4 \cite{Geant4} with the Physics List \texttt{FTFP\_BERT}, and a few
examples can be found in Fig.~\ref{fig:geant4}(a-c) (MPGD not included).
The signal is characterized by two tracks sharing a common vertex inside
the detector, which is highly distinguishable from background particles
(mainly muons) with single tracks going through the whole detector, as
illustrated by Fig.~\ref{fig:geant4}(d). For low energy muons and charged
pions, they lose energy mainly via ionization, while electrons of similar
energy level lose energy substantially via bremsstrahlung. This explains
the shorter trajectory lengths of electrons in Fig.~\ref{fig:geant4}. Muons
and pions can decay into electrons and neutrinos, or interact with the
nuclei to release neutrons, at the end of the trajectory. Both neutrons
and neutrinos leave little or no energy in the detector.

It is also possible to obtain timing information from the pattern of energy
deposition in the emulsion films alone. Figure~\ref{fig:geant4_pion} shows
the path of a 200 MeV pion entering the detector along the $z$ direction,
and the energy deposited per emulsion film. It can be seen that at the
end of the track when the pion slows down, more energy loss due to ionization
is deposited. This trend helps to tell the pion's direction of travel in
the detector.

%s9 #&#
\section{Background events}

With electrons, photons and hadrons absorbed by upstream detector and beam
pipe related infrastructures, the main background particles entering the
DM detector are muons. They consist of cosmic muons and muons produced
in $e^{+}e^{-}$ collisions, which are described in the following subsections.
A summary of the contributions from different background sources is given
in Table~\ref{tab:background}.

%t2 #&#
\begin{table}
\centering
\caption{The estimated number of muons per cm$^2$ from BESIII (at $\sqrt{s}=3.78$ GeV) or cosmic sources onto the DM detector for a given integrated luminosity or period. The production cross section at the $\psi(3686)$ resonance is taken from \cite{BESIII_design}. The rate for cosmic muons is based on sea level with no shielding. }
\label{tab:background}
\begin{tabular}{cccc}
\hline
 & non-resonant & $\psi(3686)$ & cosmic \\ \hline
lumi. or period & 10 fb$^{-1}$ & 4.5 fb$^{-1}$ & 6 months  \\ \hline
muons/cm$^2$ & 20.2 & 74.4 & $0.28\times10^6$ \\
\hline
\end{tabular}
\end{table}

%s9.1 #&#
\subsection{Collider muons}

Non-resonant $\mu ^{+}\mu ^{-}$ events are constantly produced in the
$e^{+}e^{-}$ collisions, whose differential cross section is expressed
as
%
%e17 #&#
\begin{align}
\label{equ:collider_muon}
\frac{d\sigma }{d\cos \theta } = \frac{\pi \alpha ^{2}}{2s} \left (1+
\cos ^{2}\theta \right ).
\end{align}
On the other hand, $\mu ^{+}\mu ^{-}$ events can be also produced from
hadron resonances such as $J/\psi $ and $\psi (2S)$. Because they are vectors,
their differential cross sections have the same angular dependence as in
Eq.~\ref{equ:collider_muon}. These collider muons have very small efficiency
of acceptance in the DM detector, and contribute far less background events
than the cosmic source, as shown in Table~\ref{tab:background}.

%s9.2 #&#
\subsection{Cosmic muons}

If the proposed DM detector is built above ground (BESIII) or near ground
(Belle II), comic-ray muons could be a main background. At sea level, assuming
a flat Earth, the muon flux $\Phi $ can be approximately defined as
%
%e18 #&#
\begin{align}
\label{equ:cosmic_1}
\Phi (\theta )=I_{0}\cos ^{n-1}\theta ,
\end{align}
where the zenith angle $\theta $ is the angle between the muon's momentum
and the normal of the Earth's surface, and $I_{0}$ is the vertical ($
\theta =0$) muon flux integrated over energy. With parameters
$n\approx 3$ and $I_{0}\approx 85.6\pm 2.4$ m$^{-2}$s$^{-1}$sr$^{-1}$,
obtained by fitting the previous cosmic-ray muon measurements, the muon
flux roughly follows a $\cos ^{2}\theta $ distribution
\cite{CosmicBkg,CosmicBkg2}. The integrated muon flux can be obtained as
%
%e19 #&#
\begin{align}
\label{equ:cosmic_2}
\Phi = \int _{\theta =0}^{\pi /2} \int _{\phi =-\pi }^{\pi } I_{0}\cos ^{2}
\theta d\cos \theta d\phi = 179.3~\text{Hz/m}^{2}.
\end{align}

To control the track occupancy under $10^{6}$/cm$^{2}$
\cite{occupancy_limit} for the post-processing, the emulsion films have
to be replaced before this limit is reached. With the estimated rate in
Eq.~\ref{equ:cosmic_2}, after 6 months' data taking, the cosmic occupancy
will reach $0.28\times 10^{6}$/cm$^{2}$. Therefore, replacement of the
emulsion films in every 6 months is good enough. The Belle II detector
is placed 11 meters underground in a pit. The underground rock can help
shield the cosmic muons, making this background less serious than for BESIII.
Bremsstrahlung photons from muons will give rise to EM showers containing
collimated and soft $e^{+}e^{-}$ pairs, whose topology can be well distinguished
from the signal. An emulsion detector track reconstruction algorithm is
needed to perform a detailed study. On the other hand, the MPGD+scintillator
design can also significantly reject such kind of tracks, whose time coincides
with the cosmic muons hitting the outside scintillator.

%s9.3 #&#
\subsection{$K^{0}_{L}$ and beam backgrounds}

%f13 #&#
\begin{figure}
\centering
\includegraphics[width=0.40\textwidth]{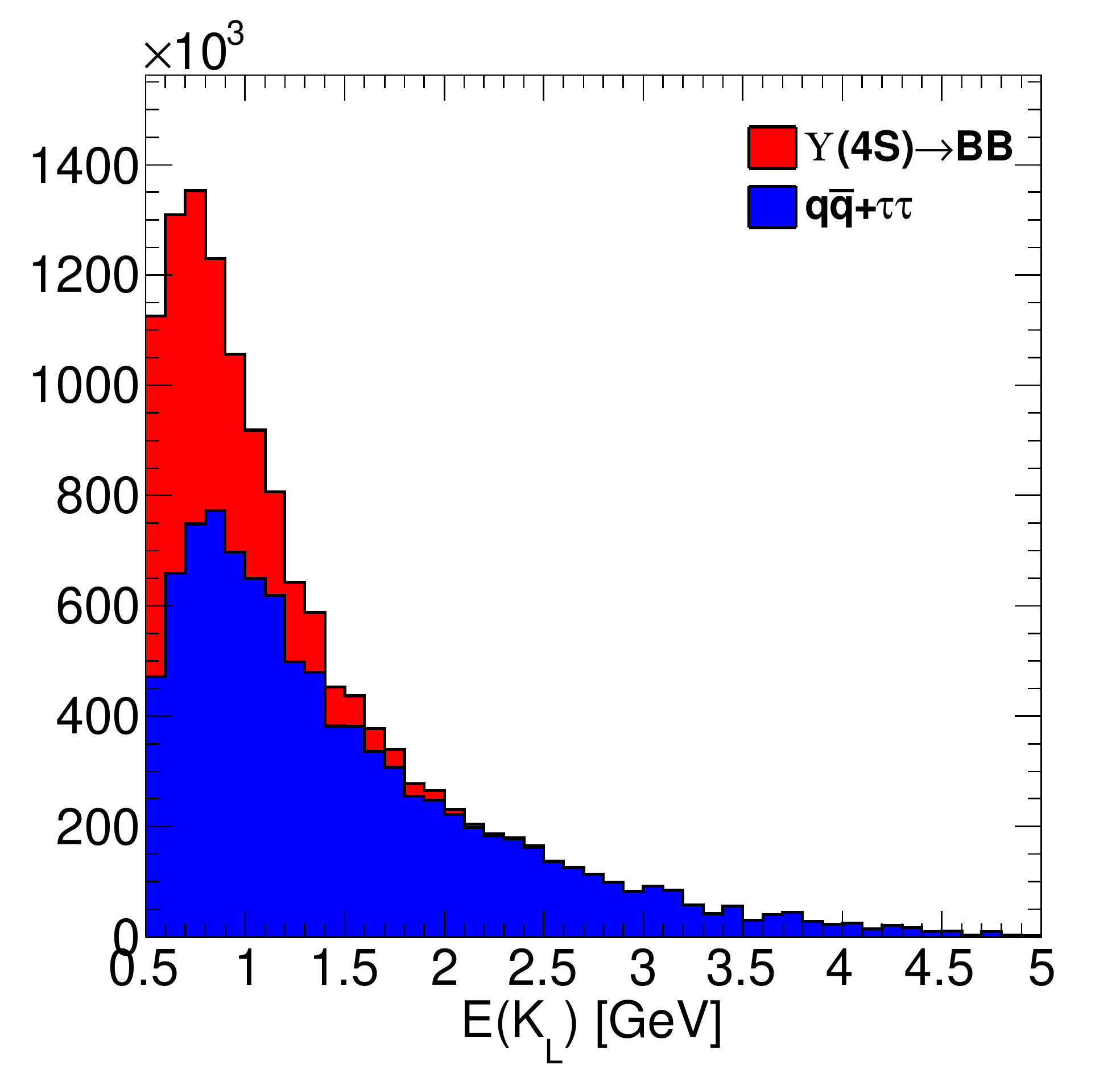}
\caption{The $K^{0}_{L}$ energy spectrum entering the proposed DM detector at
Belle II, estimated with 50 ab$^{-1}$.}
\label{fig:comment10}
\end{figure}

We simulated the Belle II production of $K^{0}_{L}$ with \texttt{EvtGen}
\cite{EvtGen} for $K^{0}_{L}$'s from $B$ meson decay, and with \texttt{KKMC}
\cite{KKMC1,KKMC2} from continuum ($u/d/s/c$) and
$\tau ^{+}\tau ^{-}$ decay. The total cross section for
$\Upsilon (4S)$ (continuum and $\tau \tau $) is 1.110 nb
\cite{Belle2book} (4.512 nb). About 69.4\% (26.0\%) events in the inclusive
$\Upsilon (4S)$ (continuum and $\tau \tau $) sample contain a
$K^{0}_{L}$. The $K^{0}_{L}$ energy spectrum entering the DM detector is
shown in Fig.~\ref{fig:comment10}, from which one can see that
$K^{0}_{L}$ from $B$ decay is much softer than from the continuum.

%f14 #&#
\begin{figure*}
\centering
\includegraphics[width=0.35\textwidth]{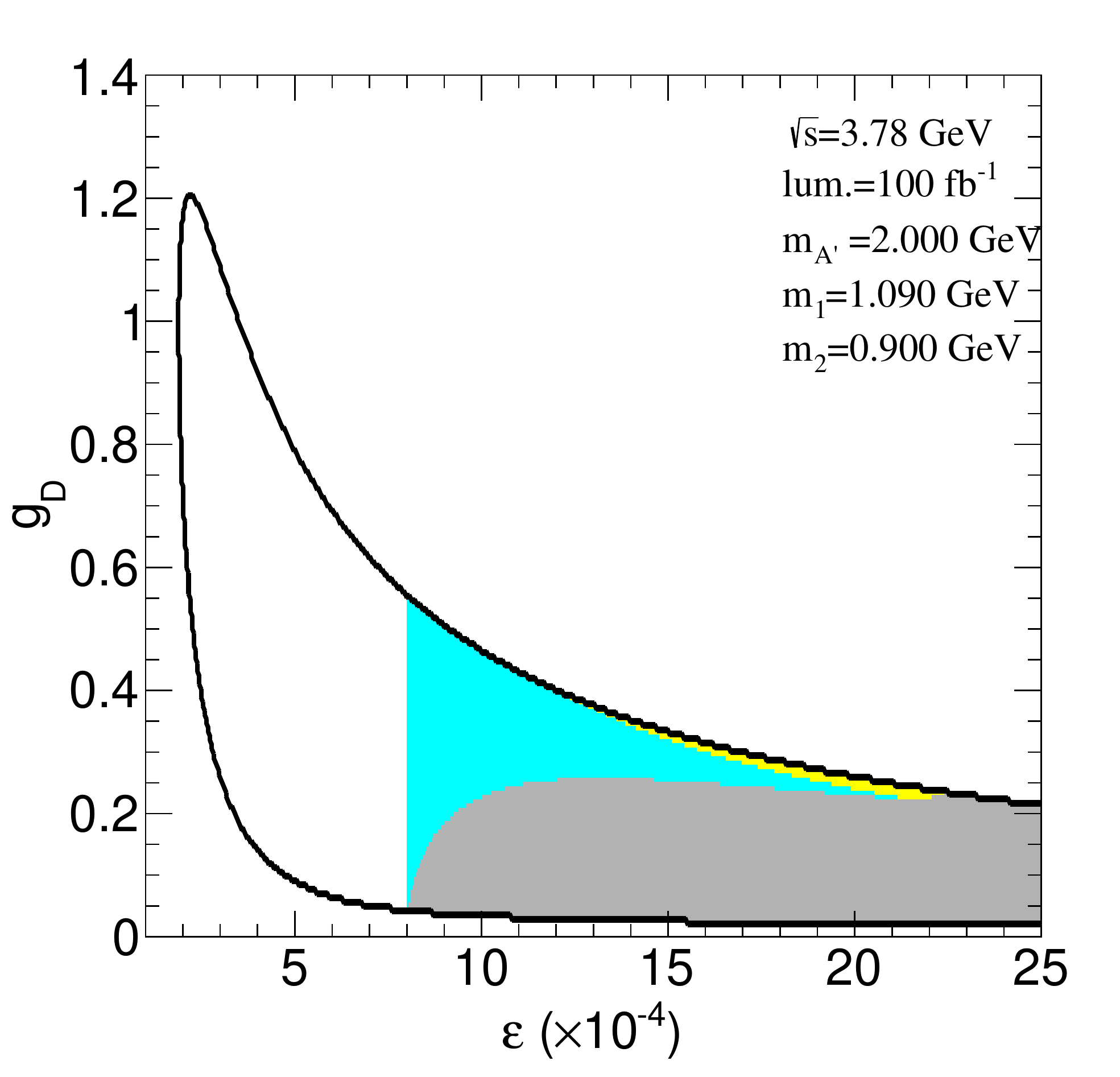} \hspace{1.5em}
\includegraphics[width=0.35\textwidth]{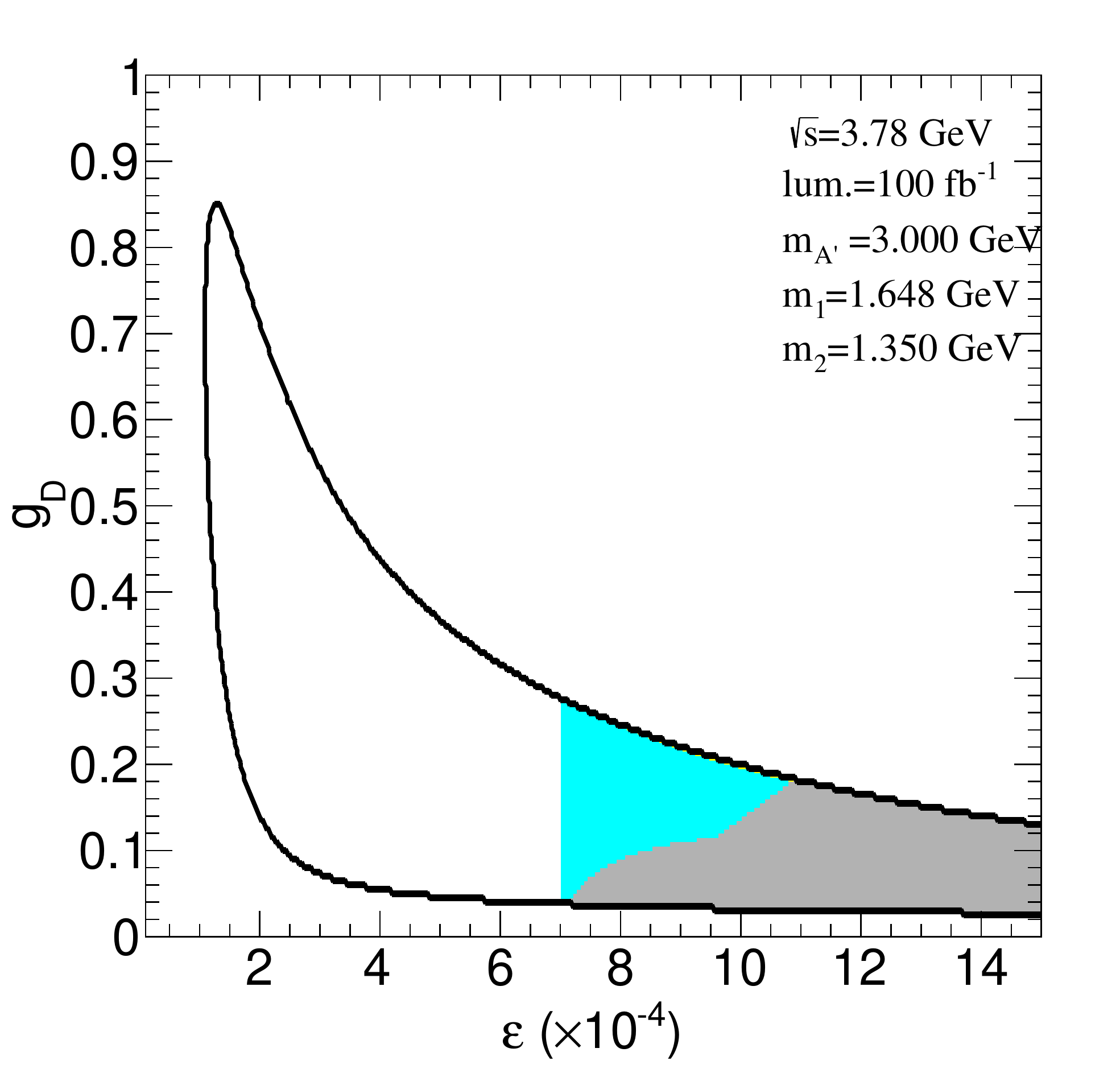} 
\put(-295, 140){\textbf{(a)}}
\put(-85, 140){\textbf{(b)}} \vspace{0.15cm} \\
\includegraphics[width=0.35\textwidth]{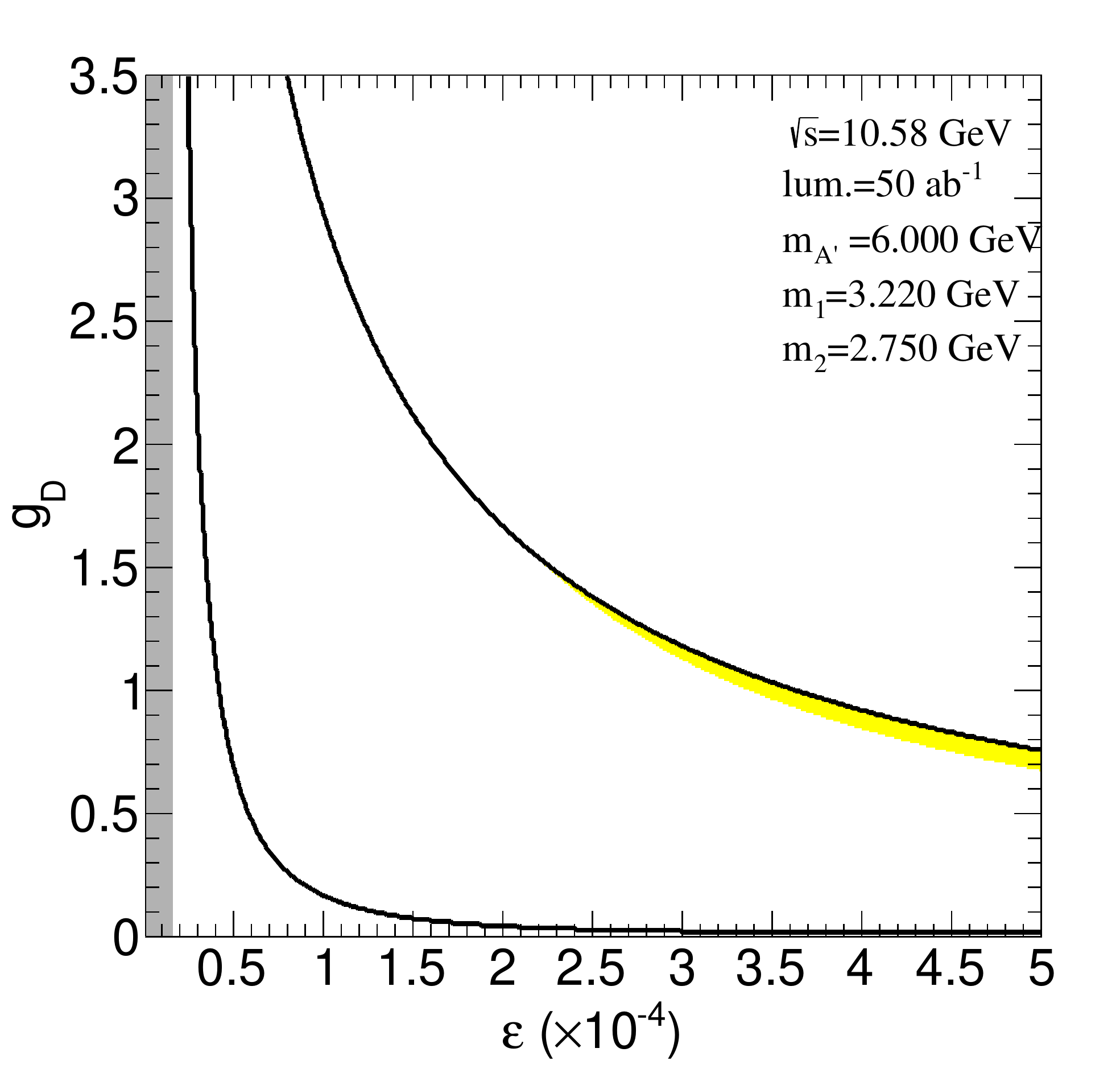} \hspace{1.5em}
\includegraphics[width=0.35\textwidth]{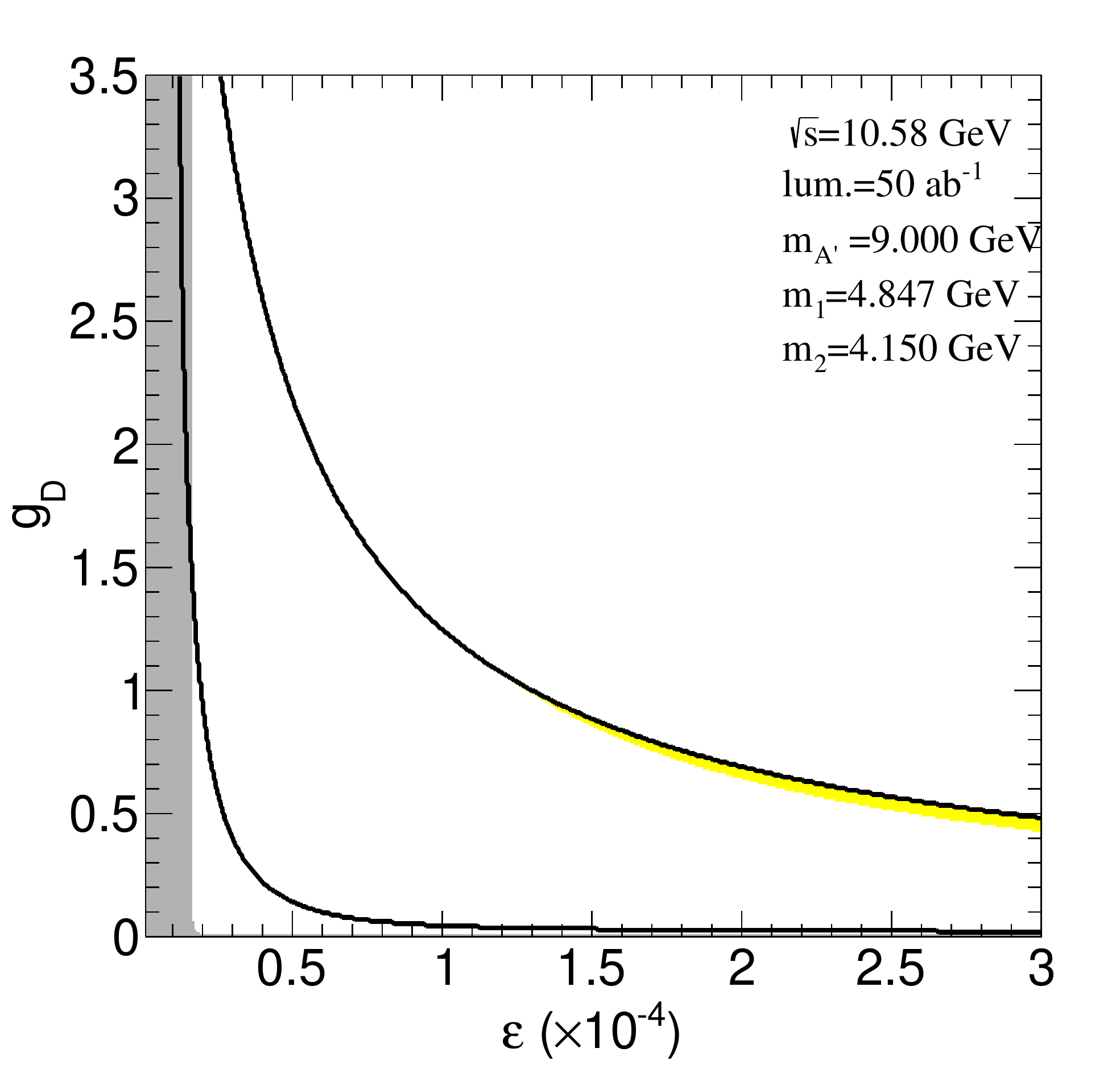}
\put(-295, 140){\textbf{(c)}}
\put(-85, 140){\textbf{(d)}}
\caption{The 90\protect\% CL exclusion regions assuming zero background in
$\epsilon $ vs. $g_{D}$ with two forward DM detectors placed on each side of
BESIII (a,~b) and Belle II (c,~d), for the chosen model parameters ($A'$ is
on-shell) indicated in the plots. The gray areas in (a,~b) have been excluded by
BaBar, and those in (c,~d) are excluded by the relic density constraint. The cyan
region could have otherwise been excluded by \cite{BaBar_AtoInv}, but re-opens for
exploration in the new model. The yellow area has an average $\phi _{1}$ flight
length $d<1$ m, which is the best suitable region for the displaced vertex plus
photon search proposed in \cite{monophoton_plus_ee}.}
\label{fig:sens1}
\end{figure*}

To block these $K^{0}_{L}$'s from IP, we propose putting a 1.3 m long of
tungsten slab in front of the forward DM detector, fully covering the detector
cross section. We then simulated the passage of $K^{0}_{L}$ through the
tungsten shield with Geant4, assuming the DM detector is 7 m away from
the IP. We further assume a 30 cm of shielding
material (iron) at about 2 m from the IP due to the beam focusing magnet
iron, the cryostat and the magnetic return yoke iron.\footnote{Exact details to
be confirmed with the experiment.} In summary, with this simulation setup
and the energy spectrum in Fig.~\ref{fig:comment10}, we found that the
probability for a $K^{0}_{L}$ to arrive at the DM detector is
$4\times 10^{-8}$. As a result, we expect $\sim $0.5 $K^{0}_{L}$ in the
DM detector coming from the IP with 50 ab$^{-1}$ of data. For the central DM detector,
since the magnetic flux return yoke is thick and can already provide a
large amount of shielding for $K^{0}_{L}$, a shorter tungsten shielding
will be sufficient.

We also simulated cosmic muons (consists of 45\% $\mu ^{-}$ and 55\%
$\mu ^{+}$) passing through 11 m of rock (whose density is $\sim $2.5 g/cm$^{3}$)
for the Belle II environment, and found that there are $\sim $12
$K^{0}_{L}$'s from cosmic muons' photonuclear interaction with the upstream
rock material passing through the DM detector per year. Most of them will
interact hadronically within the DM detector volume, instead of decay ($
\lesssim $1 \%). Therefore, the double detector setup described in Sec.~\ref{sec:det} becomes efficient to control these background events. In
addition, tracks induced by this kind of $K^{0}_{L}$'s, as well as other
neutral hadrons, can be tagged by the MPGD+scintillator design, since they
happen at the same time when the cosmic muons pass through the detector.

Beam backgrounds, which include the Touschek effect, beam-gas scattering,
synchrotron radiation and radiative Bhabha processes, can also affect the
detector. The Touschek effect and beam-gas scattering will cause beam particles
deviate from their nominal energy and hit the beam pipe walls or collimators.
Majority of the gammas from the developed EM showers are soft and can be
well shielded by the lead in Fig.~\ref{fig:emulsion_MPGD}(b). For example,
a 10 cm thick lead can block most photons with energy below a few GeV.
Again, the MPGD+scintillator design works against these backgrounds as
well. The synchrotron radiation is at tens of keV level, and hence not
a big background source. The gamma from radiative Bhabha process can hit
the magnet iron and produce a lot of neutrons. These neutrons can be well
shielded by the same tungsten slab used to shield $K^{0}_{L}$'s. However, a detailed simulation of these effects needs a good knowledge
from the detector and accelerator, which is out of the scope of this work.
Finally, we point out an interesting scenario where multiple new long-lived particles exist, and an additional secondary production of these particles may  take place in the shielding material right in front of the detector, as described in \cite{Sebastion}. Our detector setup is also sensitive to this scenario.

%s10 #&#
\section{Signal sensitivity}

The expected number of signal events in the DM detector for an on-shell
dark photon can be calculated as:
%
%e20 #&#
\begin{align}
\label{equ:yield}
N_{sig} = \mathcal{L} \sigma (e^{+}e^{-}\to \gamma A') \text{BR}(A'
\to \phi _{1}\phi _{2}) \epsilon _{A} \left ( e^{-\frac{a}{d}} - e^{-
\frac{a+L}{d}} \right ),
\end{align}
where $\mathcal{L}$ is the integrated luminosity of data, $L=1.32$ m,
$d$ is calculated as in Eq.~\ref{equ:phi1_d}, and $a=6$ m (7 m) for BESIII
(Belle II). For an off-shell $A'$, it is calculated instead by
%
%e21 #&#
\begin{align}
\label{equ:yield_off}
N_{sig} = \mathcal{L} \sigma (e^{+}e^{-}\to A^{\prime *}\to \phi _{1}
\phi _{2}) \epsilon _{off} \left ( e^{-\frac{a}{d}} - e^{-
\frac{a+L}{d}} \right ),
\end{align}
where $a=4$ m for Belle II.

%f15 #&#
\begin{figure*}
\centering
\includegraphics[width=0.35\textwidth]{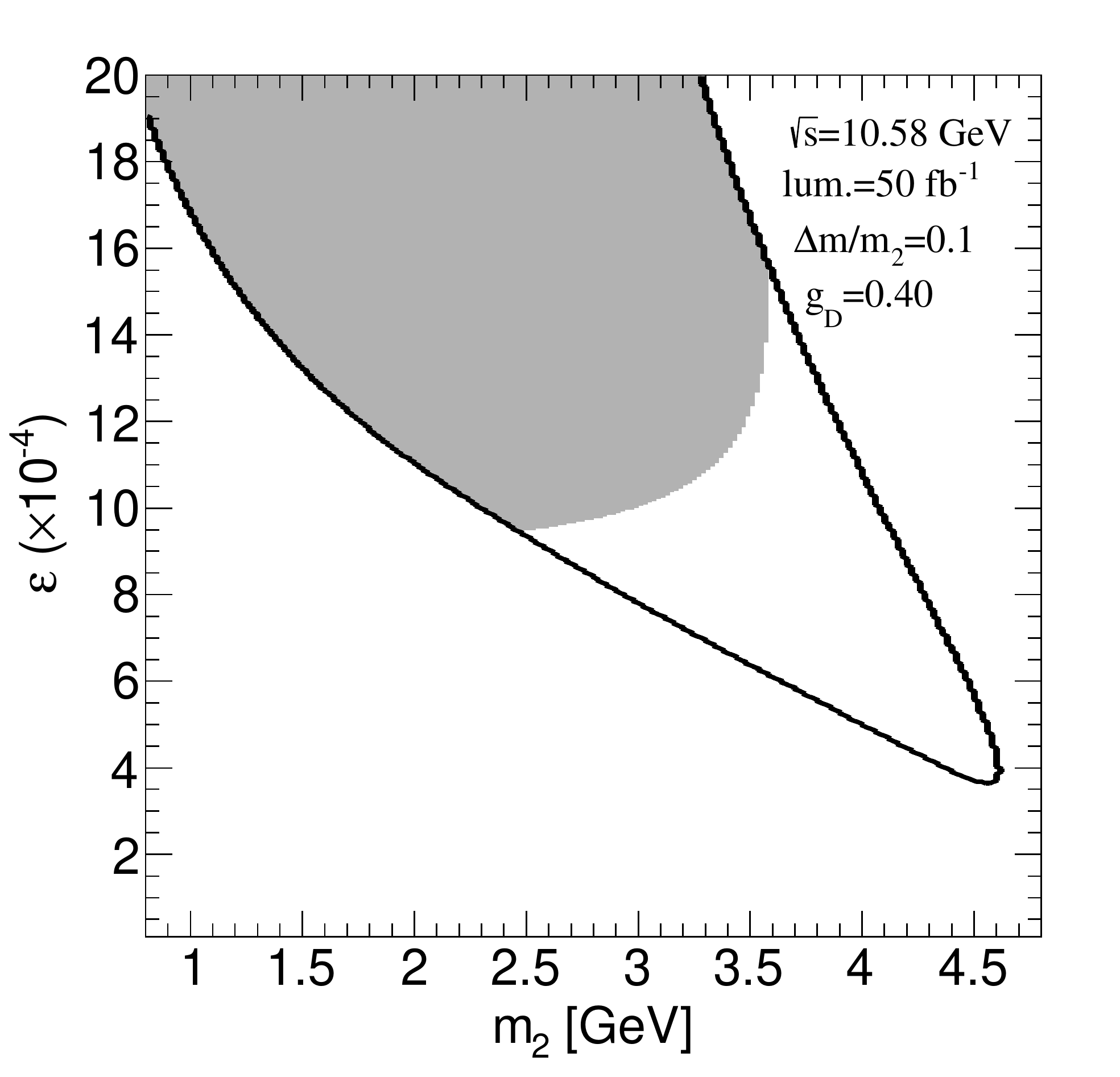} \hspace{1.5em}
\includegraphics[width=0.35\textwidth]{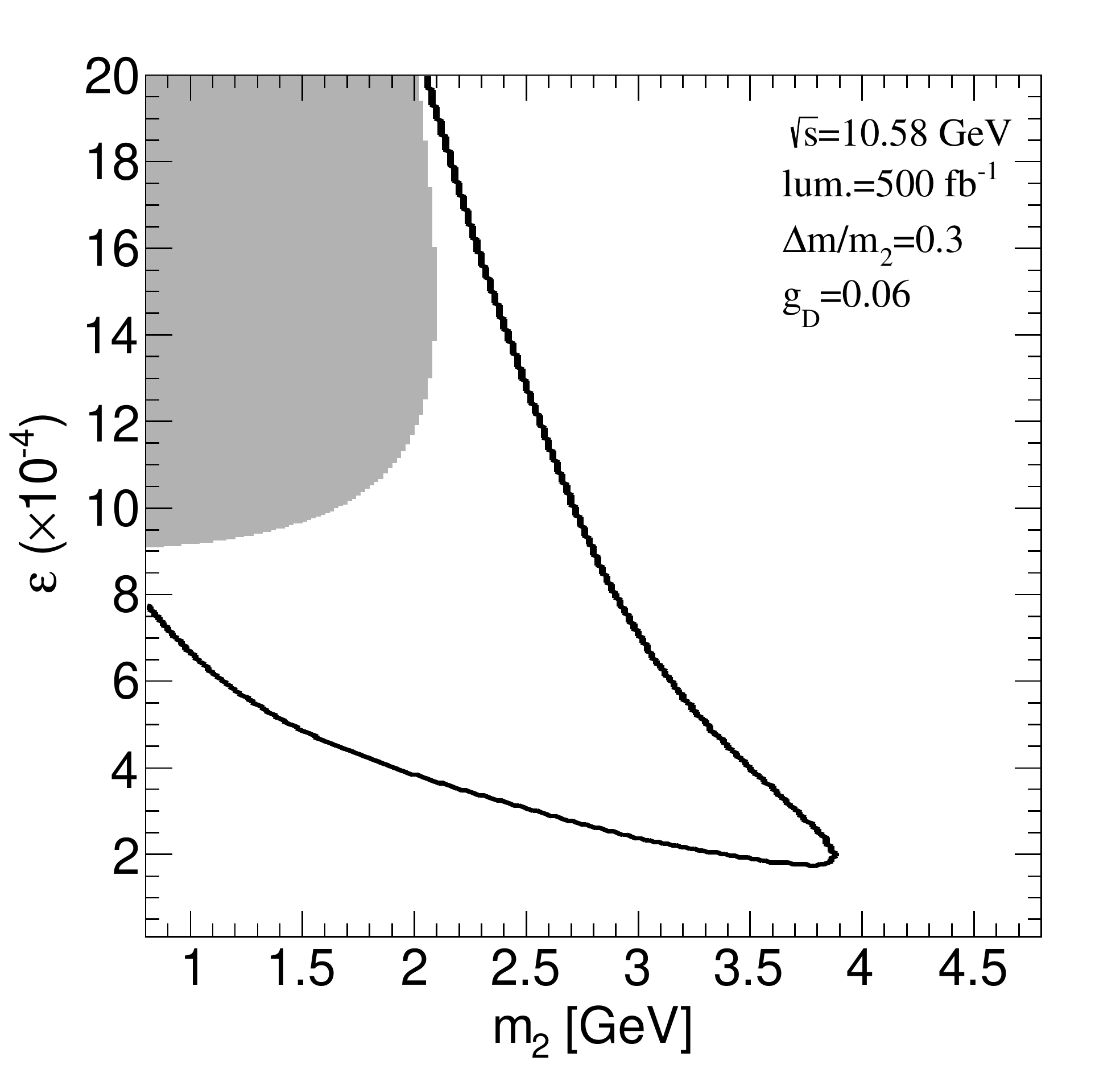} 
\put(-295, 140){\textbf{(a)}}
\put(-85, 140){\textbf{(b)}} \vspace{0.15cm} \\
\includegraphics[width=0.35\textwidth]{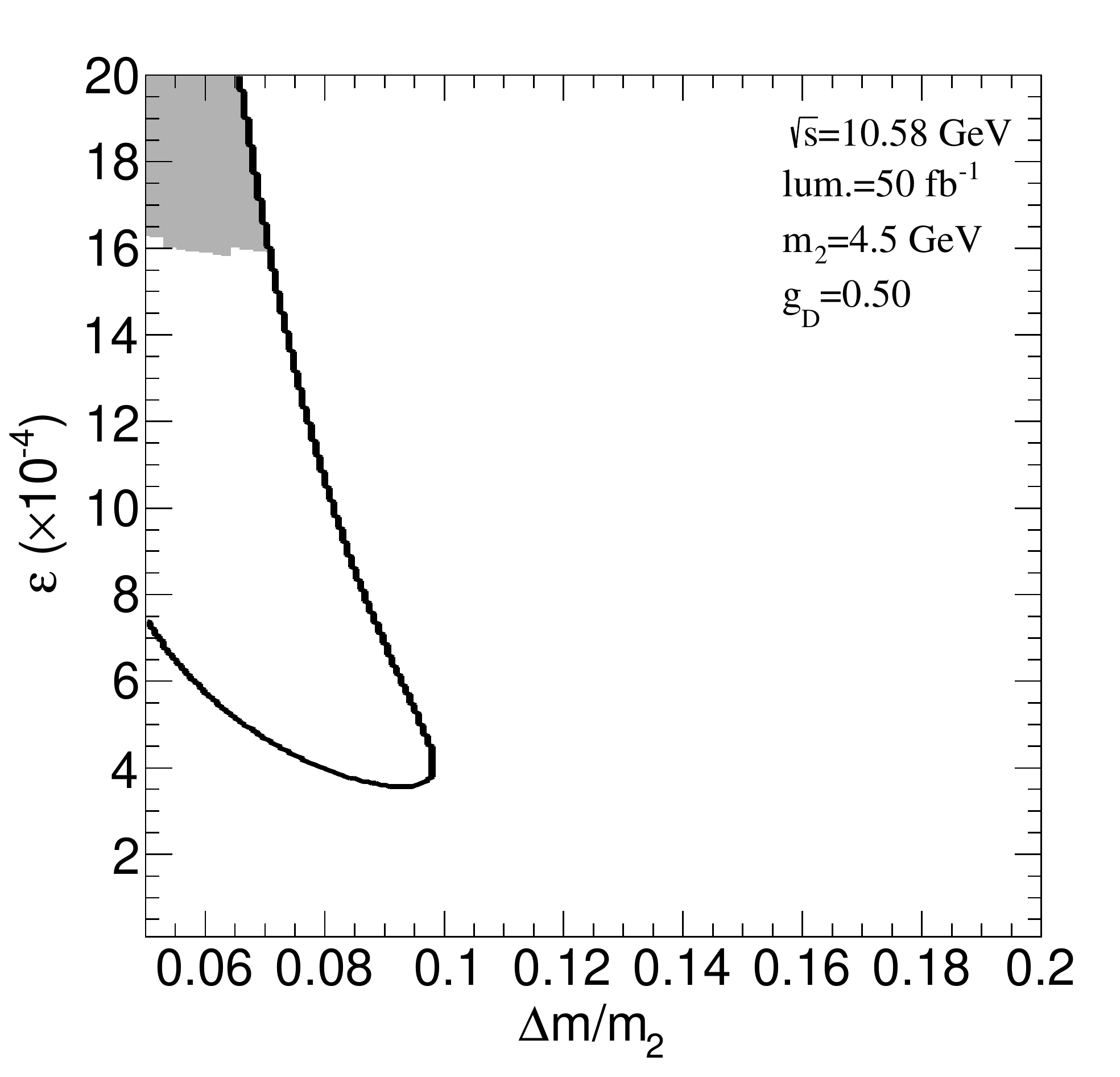} \hspace{1.5em}
\includegraphics[width=0.35\textwidth]{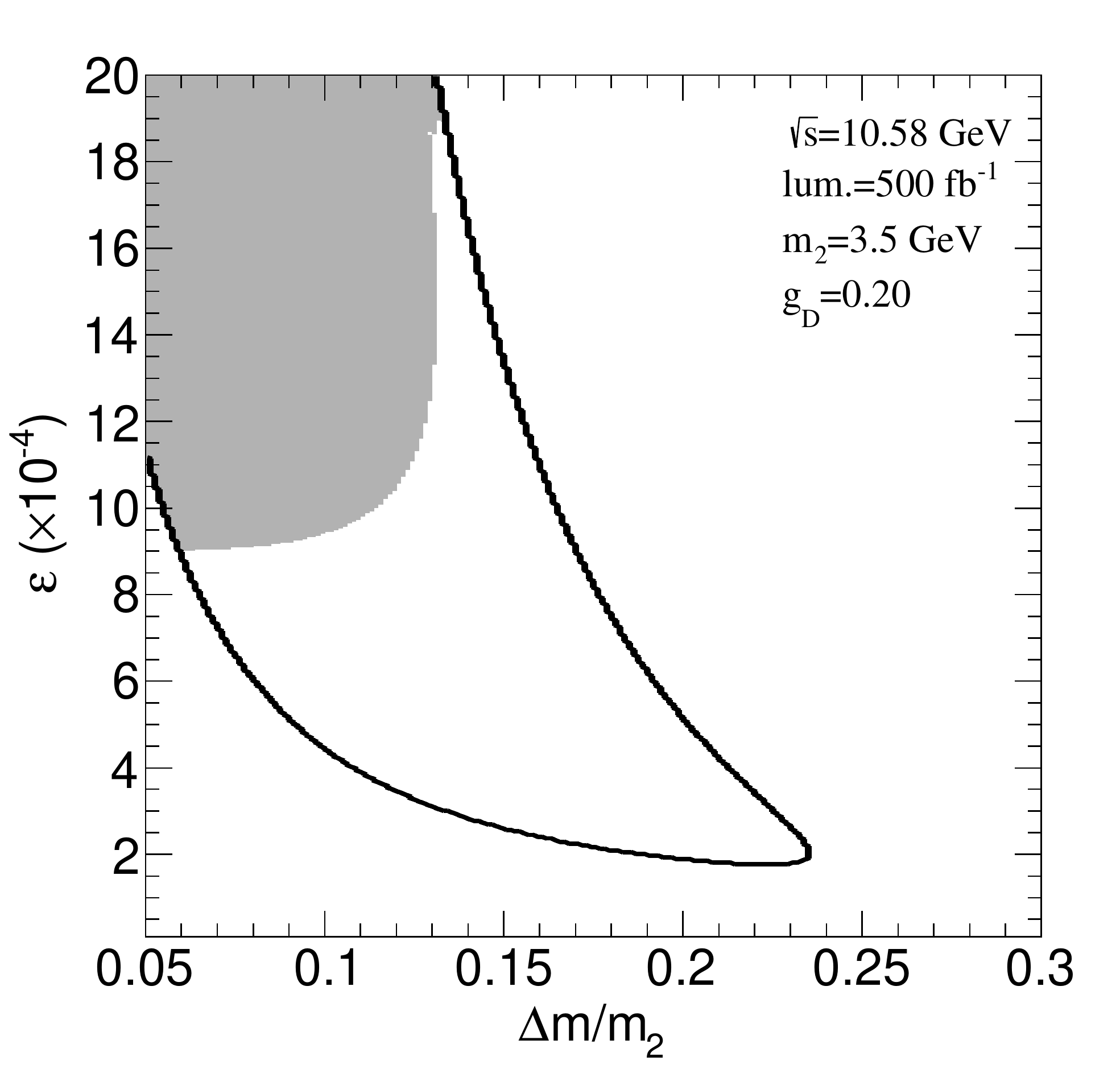}
\put(-295, 140){\textbf{(c)}}
\put(-85, 140){\textbf{(d)}}
\caption{The 90\protect\% CL exclusion regions assuming zero background in
$\epsilon $ vs. $m_{2}$ (a,~b) and $\epsilon $ vs. $\Delta m/m_{2}$ (c,~d) with
two forward DM detectors placed on each side of Belle II, for the chosen model
parameters ($A'$ is on-shell) indicated in the plots. The gray areas in these
plots have been excluded by BaBar. An optimal detector acceptance of
0.9\protect\% is assumed, as explained in the text.}
\label{fig:sens2}\vspace{6pt}
\end{figure*}

In the case of an on-shell $A'$, the parameter spaces in $\epsilon $ vs.
$g_{D}$ for some specific choices of model parameters with 90\% CL exclusion
for BESIII (Belle II) are shown in Fig.~\ref{fig:sens1}, based on 100 fb$^{-1}$
(50 ab$^{-1}$) of data, when zero background and 100\% signal reconstruction
efficiency are assumed. The gray area has been excluded by BaBar
$A'\to ll$ ($l=e,\mu $) \cite{BaBar_Atoll}, or
$A'\to \text{invisible}$ search \cite{BaBar_AtoInv}. When $\epsilon $ is
large and $g_{D}$ is small, BR($A'\to ll$) might be too large and would
have been excluded by \cite{BaBar_Atoll}. This is the case when
%
%e22 #&#
\begin{align}
\label{equ:cond_ll}
\epsilon > \frac{\epsilon _{1}}{ \sqrt{\text{BR}(A'\to ll)} },
\end{align}
where $\epsilon _{1}$ is the limit on $\epsilon $ taken from
\cite{BaBar_Atoll}. On the other hand, the search for an invisible
$A'$ is based on mono-photon events at BaBar. If $\phi _{1}$ decays inside
the BaBar detector with visible energy depositions, this event will not
be categorized into mono-photon any more. Considering the BaBar detector
size, we roughly require that $\phi _{1}$ has to fly by at least 2 m away
from IP for the event to be a mono-photon candidate. This means that events
satisfying the following condition would have been excluded by BaBar
$A'\to \text{invisible}$ search:
%
%e23 #&#
\begin{align}
\label{equ:cond_inv}
\epsilon > \epsilon _{2} e^{1/d},
\end{align}
where $\epsilon _{2}$ is the limit on $\epsilon $ taken from
\cite{BaBar_AtoInv}. The combination of Eq.~\ref{equ:cond_ll}-\ref{equ:cond_inv}
are excluded in gray in Fig.~\ref{fig:sens1}(a,~b) and \ref{fig:sens2}.
Figure~\ref{fig:sens1}-\ref{fig:sens2} assume two identical DM detectors
are placed on each side of the main $e^{+}e^{-}$ detector. The cyan regions
in Fig.~\ref{fig:sens1} could have otherwise been excluded by
\cite{BaBar_AtoInv}, but are allowed according to Eq.~\ref{equ:cond_inv}. The yellow regions have a $\phi _{1}$ with $d<1$ m,
which is the best suitable region for the displaced search proposed in 
\cite{monophoton_plus_ee}. Compared to BESIII, Belle II can probe higher
dark photon masses, and sensitive to $\epsilon $ values lower by an order
of magnitude. For $m_{A'}>8$ GeV where the mono-photon search is difficult
due to the radiative Bhabha background, the propose DM detector can well
cover this parameter region.

Similar plots showing the relations between $\epsilon $ and $m_{2}$, or
$\epsilon $ and $\Delta m/m_{2}$, are in Fig.~\ref{fig:sens2} for
Belle II with 50-500 fb$^{-1}$ of data. The parameters regions with
$m_{2}>1$ GeV that are hard to be covered by fixed target experiments,
can be well reached. In all these plots, an optimal detector acceptance
of 0.9\% is assumed. For the detector placed at box $A$ in Fig.~\ref{fig:prob_theta1_phi1}, this means that a relation of
$1-(m_{1}+m_{2})/m_{A'} = (0.228-0.0223 m_{A'})\cdot \exp (-0.489 m_{A'})$
(for $m_{A'}\leq 10$ GeV) should be satisfied. This relation is obtained
by a fit to different $m_{A'}$ and $m_{1,2}$ values where an optimal acceptance
is achieved. As mentioned in Sec.~\ref{sec:RD}, these parameter settings,
in which $m_{1}+m_{2}$ is very close to $m_{A'}$, are also favored by the
DM relic density constraint.

%f16 #&#
\begin{figure*}
\centering
\includegraphics[width=0.35\textwidth]{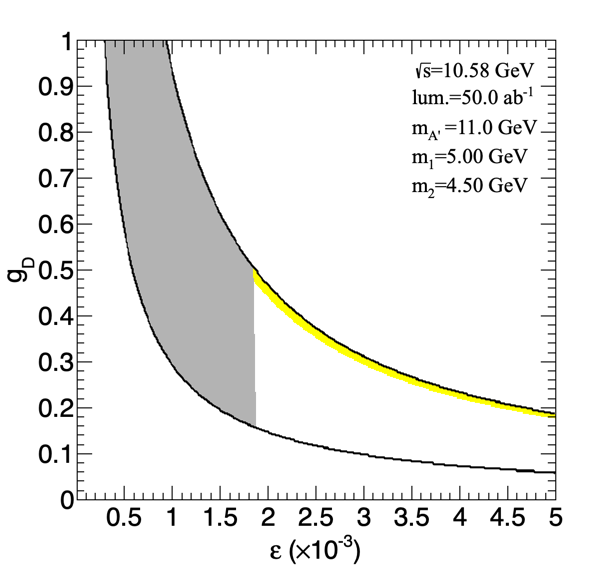} \hspace{1.5em}
\includegraphics[width=0.35\textwidth]{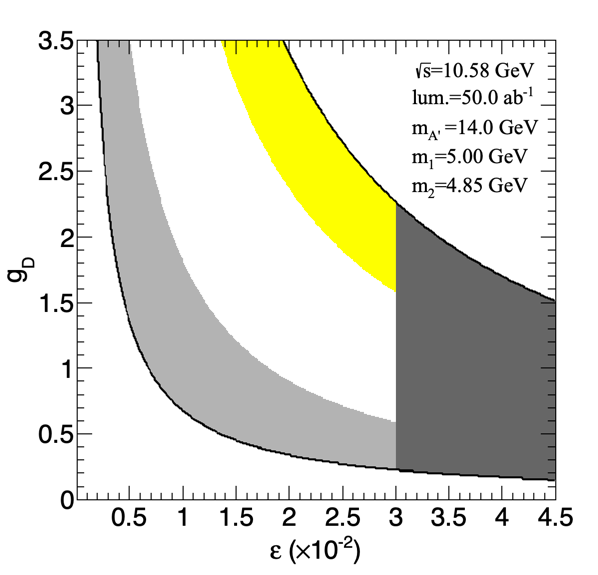} 
\caption{The 90\protect\% CL exclusion regions assuming zero background in
$\epsilon $ vs. $g_{D}$ with a single DM detector placed in the central
direction near Belle II (Fig.~\ref{fig:cad}(b)), for the chosen model
parameters ($A'$ is off-shell) indicated in the plots. The light gray areas are
excluded by the relic density constraint, and the dark gray area is excluded by
the LEP model-independent search \cite{LEP}. The yellow area has an average
$\phi _{1}$ flight length $d<1$ m.}
\label{fig:sens3}
\end{figure*}

In the case of an off-shell $A'$, the parameter spaces in
$\epsilon $ vs. $g_{D}$ for some specific choices of model parameters with
90\% CL exclusion for Belle II are shown in Fig.~\ref{fig:sens3}, when
zero background and 100\% signal reconstruction efficiency are assumed
again. In this figure, light (dark) gray areas are excluded by the relic
density constraint (LEP model-independent search \cite{LEP}).

%s11 #&#
\section{Conclusion}

If the dark photon is a mediator between the SM and dark sector, it can
be searched for directly at the current or future $e^{+}e^{-}$ colliders.
If there is a splitting in the DM mass, the heavier DM decay can be recorded
by a dedicated new small DM detector around the $e^{+}e^{-}$ colliders.
Such a DM detector can be placed in both forward and central directions,
depending on whether the dark photon is on-shell. This new detector can
search for parameter spaces not reached by previous experiments with the
new BESIII and Belle II data, and can also re-open some parameter spaces
that have been excluded by previous data derived with simple assumptions.
Such a detector around BESIII can probe the dark photon coupling parameter
$\epsilon $ down to ${\sim}10^{-4}$, and if it is around Belle II, can
further reach down to ${\sim}10^{-5}$. For an off-shell dark photon,
$\epsilon $ in the range between $10^{-3}$ and $10^{-2}$ can be excluded.
It will open a new window for new physics searches related to a dark\break sector.

\section*{Acknowledgments}
X. Chen and Z. Hu are supported by Tsinghua University Initiative Scientific Research Program. Y. Wu is supported by the Natural Sciences
 and Engineering Research Council of Canada (NSERC).

\bibliographystyle{elsarticle-num}
\bibliography{references}

\begin{thebibliography}{10}
\expandafter\ifx\csname url\endcsname\relax
  \def\url#1{\texttt{#1}}\fi
\expandafter\ifx\csname urlprefix\endcsname\relax\def\urlprefix{URL }\fi
\expandafter\ifx\csname href\endcsname\relax
  \def\href#1#2{#2} \def\path#1{#1}\fi

\bibitem{BeamDump}
{S. Andreas, C. Niebuhr and A. Ringwald}, {New limits on hidden photons from
  past electron beam dumps}, Phys. Rev. D86 (2012) 095019.
\newblock \href {http://arxiv.org/abs/1209.6083} {\path{arXiv:1209.6083}},
  \href {https://doi.org/10.1103/PhysRevD.86.095019}
  {\path{doi:10.1103/PhysRevD.86.095019}}.

\bibitem{BaBar_Atoll}
{J.P. Lees et al., BaBar Collaboration}, {Search for a dark photon in $e^+e^-$
  collisions at BABAR}, Phys. Rev. Lett 113 (2014) 201801.
\newblock \href {http://arxiv.org/abs/1406.2980} {\path{arXiv:1406.2980}},
  \href {https://doi.org/10.1103/PhysRevLett.113.201801}
  {\path{doi:10.1103/PhysRevLett.113.201801}}.

\bibitem{BaBar_AtoInv}
{J.P. Lees et al., BaBar Collaboration}, Search for invisible decays of a dark
  photon produced in $e^+e^-$ collisions at babar, Phys. Rev. Lett. 119 (2017)
  131804.
\newblock \href {http://arxiv.org/abs/1702.03327} {\path{arXiv:1702.03327}},
  \href {https://doi.org/10.1103/PhysRevLett.119.131804}
  {\path{doi:10.1103/PhysRevLett.119.131804}}.

\bibitem{XENON1T}
{E. Aprile et al., XENON Collaboration}, {Dark Matter Search Results from a One
  Tonne$\times$Year Exposure of XENON1T}, Phys. Rev. Lett. 121 (2018) 111302.
\newblock \href {http://arxiv.org/abs/1805.12562} {\path{arXiv:1805.12562}},
  \href {https://doi.org/10.1103/PhysRevLett.121.111302}
  {\path{doi:10.1103/PhysRevLett.121.111302}}.

\bibitem{L_DM_1}
{K. Schmidt-Hoberg et al.}, {Constraints on light mediators: confronting dark
  matter searches with B physics}, Physi. Lett. B727 (2013) 506.
\newblock \href {http://arxiv.org/abs/1310.6752} {\path{arXiv:1310.6752}},
  \href {https://doi.org/10.1016/j.physletb.2013.11.015}
  {\path{doi:10.1016/j.physletb.2013.11.015}}.

\bibitem{L_DM_2}
{G. Krnjaic}, {Probing Light Thermal Dark-Matter With a Higgs Portal Mediator},
  Phys. Rev. D94 (2016) 073009.
\newblock \href {http://arxiv.org/abs/1512.04119} {\path{arXiv:1512.04119}},
  \href {https://doi.org/10.1103/PhysRevD.94.073009}
  {\path{doi:10.1103/PhysRevD.94.073009}}.

\bibitem{L_DM_3}
{J. Beacham et al.}, {Physics Beyond Colliders at CERN: Beyond the Standard
  Model Working Group Report}, J. Phys. G: Nucl. Part. Phys. 47 (2019) 1.
\newblock \href {http://arxiv.org/abs/1901.09966} {\path{arXiv:1901.09966}},
  \href {https://doi.org/10.1088/1361-6471/ab4cd2}
  {\path{doi:10.1088/1361-6471/ab4cd2}}.

\bibitem{L_DM_4}
{A. Filimonova et al.}, {Probing dark sectors with long-lived particles at
  BELLE II}, Phys. Rev. D101 (2020) 095006.
\newblock \href {http://arxiv.org/abs/1911.03490} {\path{arXiv:1911.03490}},
  \href {https://doi.org/10.1103/PhysRevD.101.095006}
  {\path{doi:10.1103/PhysRevD.101.095006}}.

\bibitem{RD}
{N. Aghanim et al., Planck Collaboration}, {Planck 2018 results. VI.
  Cosmological parameters}, A\&A A6 (2020) 641.
\newblock \href {http://arxiv.org/abs/1807.06209} {\path{arXiv:1807.06209}},
  \href {https://doi.org/10.1051/0004-6361/201833910}
  {\path{doi:10.1051/0004-6361/201833910}}.

\bibitem{Excited_DM_1}
{D. Smith and N. Weiner}, {Inelastic Dark Matter}, Phys. Rev. D64 (2001)
  043502.
\newblock \href {http://arxiv.org/abs/hep-ph/0101138}
  {\path{arXiv:hep-ph/0101138}}, \href
  {https://doi.org/10.1103/PhysRevD.64.043502}
  {\path{doi:10.1103/PhysRevD.64.043502}}.

\bibitem{Excited_DM}
{E. Izaguirre et al.}, {Discovering Inelastic Thermal-Relic Dark Matter at
  Colliders}, Phys. Rev. D93 (2016) 063523.
\newblock \href {http://arxiv.org/abs/1508.03050} {\path{arXiv:1508.03050}},
  \href {https://doi.org/10.1103/PhysRevD.93.063523}
  {\path{doi:10.1103/PhysRevD.93.063523}}.

\bibitem{monophoton_plus_ee}
{M. Duerr et al.}, {Invisible and displaced dark matter signatures at Belle
  II}, JHEP 02 (2020) 039.
\newblock \href {http://arxiv.org/abs/1911.03176} {\path{arXiv:1911.03176}},
  \href {https://doi.org/10.1007/JHEP02(2020)039}
  {\path{doi:10.1007/JHEP02(2020)039}}.

\bibitem{Inelastic_had}
{A. Berlin and F. Kling}, {Inelastic Dark Matter at the LHC Lifetime
  Frontier:ATLAS, CMS, LHCb, CODEX-b, FASER, and MATHUSLA}, Phys. Rev. D99
  (2019) 015021.
\newblock \href {http://arxiv.org/abs/1810.01879} {\path{arXiv:1810.01879}},
  \href {https://doi.org/10.1103/PhysRevD.99.015021}
  {\path{doi:10.1103/PhysRevD.99.015021}}.

\bibitem{BeamDump_2}
{E. Izaguirre, Y. Kahn, G. Krnjaic and M. Moschella}, {Testing Light Dark
  Matter Coannihilation With Fixed-Target Experiments}, Phys. Rev. D96 (2017)
  055007.
\newblock \href {http://arxiv.org/abs/1703.06881} {\path{arXiv:1703.06881}},
  \href {https://doi.org/10.1103/PhysRevD.96.055007}
  {\path{doi:10.1103/PhysRevD.96.055007}}.

\bibitem{micrOMEGAs}
{G. Belanger et al.}, {micrOMEGAs5.0 : freeze-in}, Comput. Phys. Commun. 231
  (2018) 173.
\newblock \href {http://arxiv.org/abs/1801.03509} {\path{arXiv:1801.03509}},
  \href {https://doi.org/10.1016/j.cpc.2018.04.027}
  {\path{doi:10.1016/j.cpc.2018.04.027}}.

\bibitem{QiaoCongfeng}
{J. Jiang, H. Yang and C.-F. Qiao}, {Exploring Bosonic Mediator of Interaction
  at BESIII}, Euro. Phys. Jour. C79 (2019) 404.
\newblock \href {http://arxiv.org/abs/1810.05790} {\path{arXiv:1810.05790}},
  \href {https://doi.org/10.1140/epjc/s10052-019-6912-3}
  {\path{doi:10.1140/epjc/s10052-019-6912-3}}.

\bibitem{PDG}
{M. Tanabashi et al., Particle Data Group},
  \href{http://pdg.lbl.gov/2019/hadronic-xsections/hadron.html}{{Review of
  Particle Physics}}, Phys. Rev. D98 (2018) 030001.
\newblock \href {https://doi.org/10.1103/PhysRevD.98.030001}
  {\path{doi:10.1103/PhysRevD.98.030001}}.
\newline\urlprefix\url{http://pdg.lbl.gov/2019/hadronic-xsections/hadron.html}

\bibitem{Opera}
{M. Komatsu}, \href{http://cds.cern.ch/record/642973}{{OPERA experiment}},
  Nucl. Instrum. Meth. A503 (2003) 124.
\newblock \href {https://doi.org/10.1016/S0168-9002(03)00653-3}
  {\path{doi:10.1016/S0168-9002(03)00653-3}}.
\newline\urlprefix\url{http://cds.cern.ch/record/642973}

\bibitem{DsTau}
{S. Aoki et al., DsTau Collaboration}, {DsTau: study of tau neutrino production
  with 400 GeV protons from the CERN-SPS}, JHEP 01 (2020) 033.
\newblock \href {http://arxiv.org/abs/1906.03487} {\path{arXiv:1906.03487}},
  \href {https://doi.org/10.1007/JHEP01(2020)033}
  {\path{doi:10.1007/JHEP01(2020)033}}.

\bibitem{FASERnu}
{H. Abreu et al., FASER Collaboration}, {Detecting and Studying High-Energy
  Collider Neutrinos with FASER at the LHC}, Euro. Phys. Jour. C80 (2020) 61.
\newblock \href {http://arxiv.org/abs/1908.02310} {\path{arXiv:1908.02310}},
  \href {https://doi.org/10.1140/epjc/s10052-020-7631-5}
  {\path{doi:10.1140/epjc/s10052-020-7631-5}}.

\bibitem{Geant4}
{S. Agostinelli et al.}, \href{http://cds.cern.ch/record/602040}{{Geant4 — a
  simulation toolkit}}, Nucl. Instrum. Meth. A506 (2003) 250.
\newblock \href {https://doi.org/10.1016/S0168-9002(03)01368-8}
  {\path{doi:10.1016/S0168-9002(03)01368-8}}.
\newline\urlprefix\url{http://cds.cern.ch/record/602040}

\bibitem{mRWELL}
{L. Shekhtman et al.}, {Development of $\mu$-RWELL detectors for the upgrade of
  the tracking system of CMD-3 detector}, Nucl. Instrum. Meth. A936 (2019) 401.
\newblock \href {https://doi.org/10.1016/j.nima.2018.11.033}
  {\path{doi:10.1016/j.nima.2018.11.033}}.

\bibitem{BESIII_design}
{M. Ablikim et al.}, {Design and Construction of the BESIII Detector}, Nucl.
  Instrum. Meth. A614 (2010) 345.
\newblock \href {http://arxiv.org/abs/0911.4960} {\path{arXiv:0911.4960}},
  \href {https://doi.org/10.1016/j.nima.2009.12.050}
  {\path{doi:10.1016/j.nima.2009.12.050}}.

\bibitem{CosmicBkg}
{K. Altenmuller et al., KATRIN Collaboration}, {Muon-induced background in the
  KATRIN main spectrometer}, Astropart. Phys. 108 (2019) 40.
\newblock \href {http://arxiv.org/abs/1805.12173} {\path{arXiv:1805.12173}},
  \href {https://doi.org/10.1016/j.astropartphys.2019.01.003}
  {\path{doi:10.1016/j.astropartphys.2019.01.003}}.

\bibitem{CosmicBkg2}
{P. Shukla and S. Sankrith}, {Energy and angular distributions of atmospheric
  muons at the Earth}\href {http://arxiv.org/abs/1606.06907}
  {\path{arXiv:1606.06907}}.

\bibitem{occupancy_limit}
{T. Fukuda et al.}, {The analysis of interface emulsion detector for the OPERA
  experiment in JAPAN Scanning facility}, J. of Instrum. 5 (2010) 04009.
\newblock \href {https://doi.org/10.1088/1748-0221/5/04/P04009}
  {\path{doi:10.1088/1748-0221/5/04/P04009}}.

\bibitem{EvtGen}
{D. J. Lange}, {The EvtGen particle decay simulation package}, Nucl. Instrum.
  Meth. A462 (2001) 152.
\newblock \href {https://doi.org/10.1016/S0168-9002(01)00089-4}
  {\path{doi:10.1016/S0168-9002(01)00089-4}}.

\bibitem{KKMC1}
{S. Jadach, B.F.L. Ward and Z. Was}, {The precision Monte Carlo event generator
  KK for two-fermion final states in $e^+ e^-$ collisions}, Comput. Phys.
  Commun. 130 (2000) 260.
\newblock \href {http://arxiv.org/abs/hep-ph/9912214}
  {\path{arXiv:hep-ph/9912214}}, \href
  {https://doi.org/10.1016/S0010-4655(00)00048-5}
  {\path{doi:10.1016/S0010-4655(00)00048-5}}.

\bibitem{KKMC2}
{S. Jadach, B.F.L. Ward and Z. Was}, {Coherent exclusive exponentiation for
  precision Monte Carlo calculations}, Phys. Rev. D63 (2001) 113009.
\newblock \href {http://arxiv.org/abs/hep-ph/0006359}
  {\path{arXiv:hep-ph/0006359}}, \href
  {https://doi.org/10.1103/PhysRevD.63.113009}
  {\path{doi:10.1103/PhysRevD.63.113009}}.

\bibitem{Belle2book}
{E. Kou et al., Belle II Collaboration}, {The Belle II Physics Book}, Prog.
  Theor. Exp. Phys. 12 (2019) 123C01.
\newblock \href {http://arxiv.org/abs/1808.10567} {\path{arXiv:1808.10567}},
  \href {https://doi.org/10.1093/ptep/ptz106} {\path{doi:10.1093/ptep/ptz106}}.

\bibitem{Sebastion}
{K. Jodłowski, F. Kling, L. Roszkowski and S. Trojanowski}, {Extending the
  reach of FASER, MATHUSLA, and SHiP towards smaller lifetimes using secondary
  particle production}, Phys. Rev. D101 (2020) 095020.
\newblock \href {http://arxiv.org/abs/1911.11346} {\path{arXiv:1911.11346}},
  \href {https://doi.org/10.1103/PhysRevD.101.095020}
  {\path{doi:10.1103/PhysRevD.101.095020}}.

\bibitem{LEP}
{A. Hook, E. Izaguirre and J. G. Wacker}, {Model Independent Bounds on Kinetic
  Mixing}, Adv. High Energy Phys. 2011 (2011) 859762.
\newblock \href {http://arxiv.org/abs/1006.0973} {\path{arXiv:1006.0973}},
  \href {https://doi.org/10.1155/2011/859762} {\path{doi:10.1155/2011/859762}}.

\end{thebibliography}

\end{document}